\documentclass[pdflatex,sn-nature]{sn-jnl}

\usepackage{graphicx}%
\usepackage{multirow}%
\usepackage{amsmath,amssymb,amsfonts}%
\usepackage{amsthm}%
\usepackage{mathrsfs}%
\usepackage[title]{appendix}%
\usepackage{xcolor}%
\usepackage{textcomp}%
\usepackage{manyfoot}%
\usepackage{booktabs}%
\usepackage{algorithm}%
\usepackage{algorithmicx}%
\usepackage{algpseudocode}%
\usepackage{listings}%

\usepackage{gensymb}
\usepackage{ulem}
\usepackage[resetlabels]{multibib}
\newcites{si}{References}

\theoremstyle{thmstyleone}%
\theoremstyle{thmstyletwo}%

\theoremstyle{thmstylethree}%

\begin{document}

\title[Article Title]{\bf{A new collective mode in an iron-based superconductor with electronic nematicity}}


\author[1]{\fnm{Haruki} \sur{Matsumoto}}

\author[2]{\fnm{Silvia} \sur{Neri}}

\author[3]{\fnm{Tomoki} \sur{Kobayashi}}

\author[3]{\fnm{Atsutaka} \sur{Maeda}}

\author[2]{\fnm{Dirk} \sur{Manske}}

\author*[1,4]{\fnm{Ryo} \sur{Shimano}}\email{shimano@phys.s.u-tokyo.ac.jp}

\affil[1]{\orgdiv{Department of Physics}, \orgname{The University of Tokyo}, \orgaddress{\street{Hongo}, \city{Bunkyo-ku}, \postcode{113-0033}, \state{Tokyo}, \country{Japan}}}

\affil[2]{\orgname{Max Planck Institute for Solid State Research}, \orgaddress{\street{Heisenberg str. 1}, \postcode{70569}, \state{Stuttgart}, \country{Germany}}}

\affil[3]{\orgdiv{Department of Basic Science}, \orgname{The University of Tokyo}, \orgaddress{\street{Komaba}, \city{Meguro-ku}, \postcode{153-8902}, \state{Tokyo}, \country{Japan}}}

\affil[4]{\orgdiv{Cryogenic Research Center}, \orgname{The University of Tokyo}, \orgaddress{\street{Yayoi}, \city{Bunkyo-ku}, \postcode{113-0032}, \state{Tokyo}, \country{Japan}}}


\abstract{Elucidation of the symmetry and structure of order parameter(OP) is a fundamental subject in the study of superconductors. Recently, a growing number of superconducting materials have been identified that suggest additional spontaneous symmetry breakings besides the primal breaking of U(1) gauge symmetry, including time-reversal, chiral, and rotational symmetries. Observation of collective modes in those exotic superconductors is particularly important, as they provide the fingerprints of the superconducting OP. Here we investigate the collective modes in an iron-based superconductor, FeSe, a striking example of superconductivity emergent in an electronic nematic phase where the rotational symmetry of electronic degree of freedom is spontaneously broken. By using terahertz nonlinear spectroscopy technique, we discovered a collective mode resonance located substantially below the superconducting gap energy, distinct from the amplitude Higgs mode. Comparison with theoretical calculations demonstrates that the observed mode is attributed to a collective fluctuation between the $s+d$-wave-like ground state and the subleading pairing channel, which corresponds to the so-called Bardasis-Schrieffer mode but also resembles an {\it intraband} Leggett mode. Our result corroborates the multicomponent pairing channels in FeSe activated in the lower space group symmetry in the electronic nematic phase.}

\keywords{Iron-based superconductors, Collective modes, Terahertz, Electronic nematicity}



\maketitle

\section*{Main}\label{sec:intro}

Iron-chalcogenide superconductors have attracted much attention due to their captivating superconducting properties, including the superconductivity in the presence of electron nematicity \cite{doi:10.7566/JPSJ.89.102002,sym12091402, doi:10.1073/pnas.1606562113, doi:10.1073/pnas.1418994112, doi:10.1073/pnas.2110501119,bohmerNematicityNematicFluctuations2022a}, high tunability of the superconducting critical temperature ($T_{\text{c}}$) by pressure \cite{Sun2016} or intercalation \cite{PhysRevB.82.180520}, 
Cooper pairing in the BCS-BEC crossover regime \cite{doi:10.1073/pnas.1413477111}, and the topological superconductivity where the Majorana bound states are anticipated to exist in the vortex core \cite{doi:10.1126/science.aan4596}. 

In bulk FeSe, a nematic order appears around $T_{\text{s}}\sim 90\, \text{K}$ and the superconductivity appears below $T_{\text{c}}\sim 8\, \text{K}$ \cite{doi:10.1073/pnas.0807325105}. The superconducting gap in the hole pocket around the $\Gamma$ point and in the electron pocket around the M point exhibit strongly anisotropic two-fold ($C_{2}$) rotationally symmetry as revealed by scanning tunneling spectroscopy(STS) \cite{doi:10.1126/science.aal1575} and by angle-resolved photoemission spectroscopy(ARPES) \cite{PhysRevX.8.031033,PhysRevB.97.180501,Hashimoto2018}. The tetragonal ($C_{4}$) to orthorhombic ($C_{2}$) symmetry breaking in the nematic phase allows the $d$-wave pairing channel to take place originating from the repulsive interactions between the electron Fermi pockets\cite{PhysRevLett.115.026402,PhysRevLett.120.267001,PhysRevB.96.045130}. Potentially, this allows the system to develop into a superconducting state with mixed pairing channels, however, the superconducting order parameter (OP) has remained under intensive debates \cite{Islam2024}.

The fingerprints of the superconducting OP are expected to appear in the behavior of collective modes, like in the case of superfluid $^3$He where a rich variety of collective modes reflecting the exotic superfluid phases has been elucidated\cite{RevModPhys.69.645}.
In general, spontaneous breaking of U(1) gauge symmetry in superconductors leads to the appearance of two collective excitations: the Goldstone mode, the global phase mode of the OP, and the Higgs mode which corresponds to amplitude fluctuations of the OP \cite{annurev:/content/journals/10.1146/annurev-conmatphys-031119-050813}.

Whenever the OP is represented by a complex admixiture of multiple pairing channels, the collective excitation spectrum becomes richer. In particular, given an order parameter of N components, we expect N amplitude modes and N-1 phase modes, with one global phase mode lifted up to the plasma frequency due to the so-called Anderson-Higgs mechanism. These modes can potentially hybridize, some manifesting as distinct excitations within the sub-gap region \cite{PhysRevB.87.054503, Poniatowski2022}.
When the OP is consisted of a single component of, e.g., $s$-wave pairing, but a sub-dominant pairing fluctuation, e.g. of $d$-wave, is present, an exciton-like collective mode appears below the superconducting gap energy, referred to as the Bardasis-Shrieffer mode \cite{PhysRev.121.1050,PhysRevB.80.140512,PhysRevB.103.024519,Grasset2022,PhysRevB.89.245134}.

Over the last decade, the terahertz(THz) nonlinear spectroscopy technique has developed as a powerful tool to access low energy collective modes including the Higgs mode in superconductors \cite{PhysRevLett.111.057002,doi:10.1126/science.1254697, annurev:/content/journals/10.1146/annurev-conmatphys-031119-050813,vaswaniLightQuantumControl2021}. Among various THz nonlinear spectroscopy techniques, here we shed light on the third-harmonic generation (THG). THz-THG associated with the Higgs mode was initially identified in a conventional $s$-wave superconductor NbN \cite{,doi:10.1126/science.1254697}, revealing the two-photon resonance of the Higgs mode at $2\omega=2\Delta$ where $\omega$ is the frequency of the incident THz wave and $2\Delta$ is the superconducting gap. Subsequently, the THz-THG technique has been applied to cuprate SCs \cite{Chu2020,doi:10.1126/sciadv.adg9211} and to a conventional multiband superconductor $\text{MgB}_{2}$ \cite{PhysRevB.104.L140505} and $\text{FeSe}_{0.5}\text{Te}_{0.5}$ \cite{Isoyama2021} to study the behavior of collective modes. In addition to the Higgs mode, it has been theoretically shown that the BS mode gives rise to THz-THG with a resonance exhibiting around $2\omega=\omega_{\text{m}}$, where $\omega_{\text{m}}$ represents the eigenfrequency of the collective mode \cite{PhysRevB.104.144508}. 

In this paper, we applied the THz-THG spectroscopy to study the low energy collective modes with the aim of gaining insight into the structure of the superconducting OP in an epitaxially grown thin film of FeSe. 

\section*{THz-THG experiments}\label{sec:THG}

\begin{figure}[h]
\centering
\includegraphics[width=0.9\textwidth]{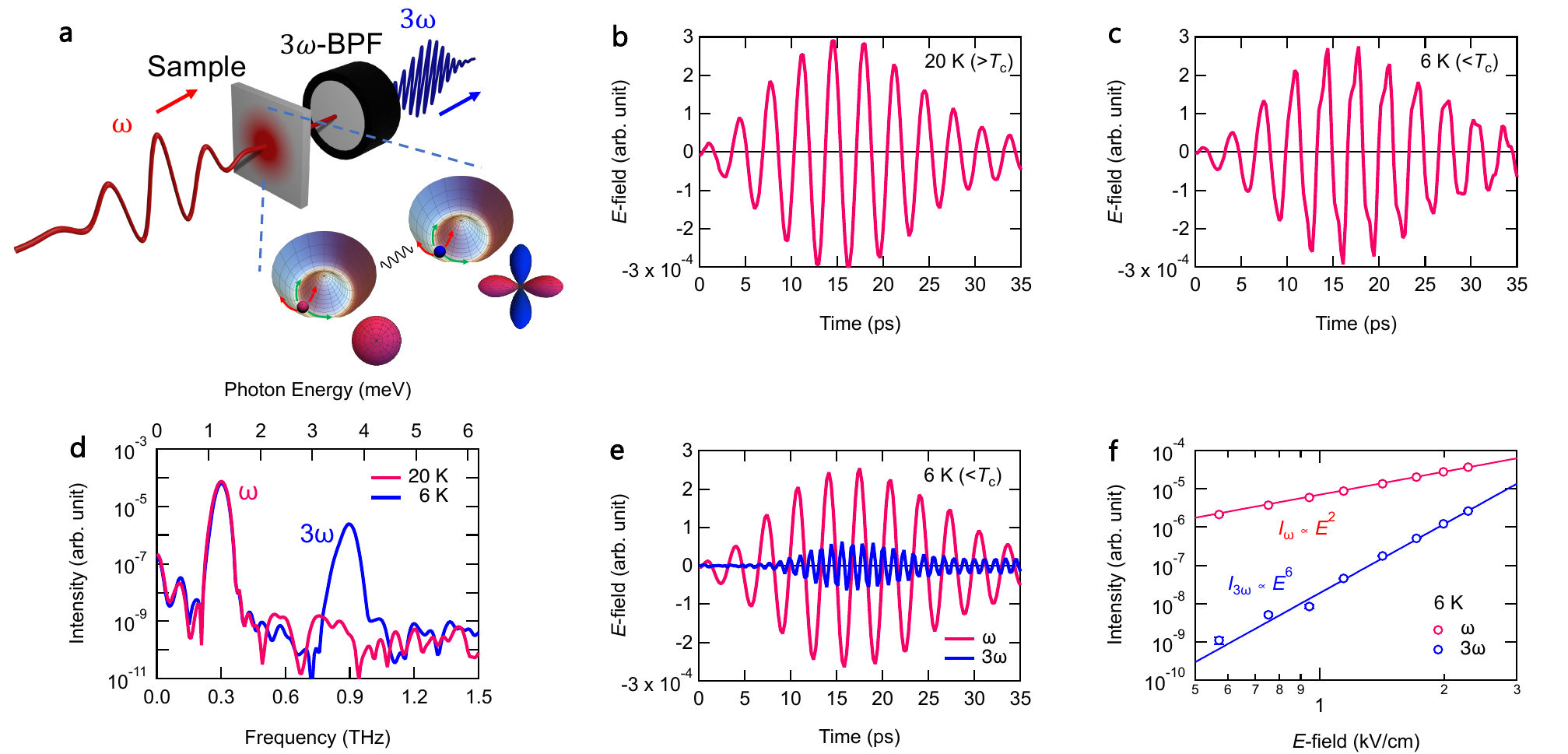}
\caption{{\bf a} Schematic experimental configuration for the THz-THG measurements. BPF : band-pass filter. {\bf b, c} Waveforms of the transmitted THz pulses above ($20\, \text{K}$) and below ($6\, \text{K}$) $T_{\text{c}}=13.8\, \text{K}$, respectively. {\bf d} Power spectra of the transmitted THz pulses above ($20\, \text{K}$) and below ($6\, \text{K}$) $T_{\text{c}}$, respectively. {\bf e} $0.3\, \text{THz}$ fundamental harmonic (FH) and $0.9\, \text{THz}$ third harmonic (TH) components extracted from (c) using $0.61\, \text{THz}$-FFT low pass and high pass filters. {\bf f} THG intensity as a function of the incident THz electric field strength.}\label{fig:1}
\end{figure}

To observe THz-THG, a narrow-band THz pulse with the center frequency $\omega$ was focused on the sample. The TH component ($3\omega$) of the transmitted pulse was measured through a bandpass filter (BPF) as schematically illustrated in Fig. \ref{fig:1}a. We used a thin film of FeSe grown by pulsed laser deposition on $\text{CaF}_{2}$ substrate (see “Methods” and Supplementary Note 1). 

Figures \ref{fig:1}b and \ref{fig:1}c show the transmitted pulses for the incident frequency of $\omega = 0.3\, \text{THz}$ above and below $T_{\text{c}} = 13.8\, \text{K}$, respectively. The corresponding power spectra are plotted in Fig. \ref{fig:1}d, showing the appearance of the $3\omega$ peak below $T_{\text{c}}$. The fundamental harmonic (FH) and TH signal are extracted from the transmitted waveform below $T_{\text{c}}$ using FFT analysis as plotted in Fig. \ref{fig:1}e. The intensities of FH ($I_{\omega}$) and TH ($I_{3\omega}$) were determined by integrating the power spectrum around $\omega$ and $3\omega$, respectively. The phases of FH ($\Phi_{\omega}$) and TH ($\Phi_{3\omega}$) were determined as those of the corresponding Fourier components at frequencies $\omega$ and $3\omega$, respectively (see Supplementary Note 4 for more details). The fluence dependence of $I_{\omega}$ and $I_{3\omega}$ is plotted in Fig. \ref{fig:1}f, ensuring the behavior of $I_{3\omega} \propto I_{\omega}^{3}$ expected for the third-order nonlinear process within the perturbative regime.

To detect the resonance of collective mode, we measured the temperature dependence of the THz-THG. Before considering THG in FeSe, we first show how the collective mode resonance appears in the temperature dependence of THG efficiency and its phase in a thin film of conventional superconductor NbN for reference, where the origin of THG is dominated by the Higgs mode \cite{annurev:/content/journals/10.1146/annurev-conmatphys-031119-050813,PhysRevB.99.224510,PhysRevResearch.2.043029} (see 'Methods' for details of the NbN sample).

\begin{figure}[h]
\centering
\includegraphics[width=0.9\textwidth]{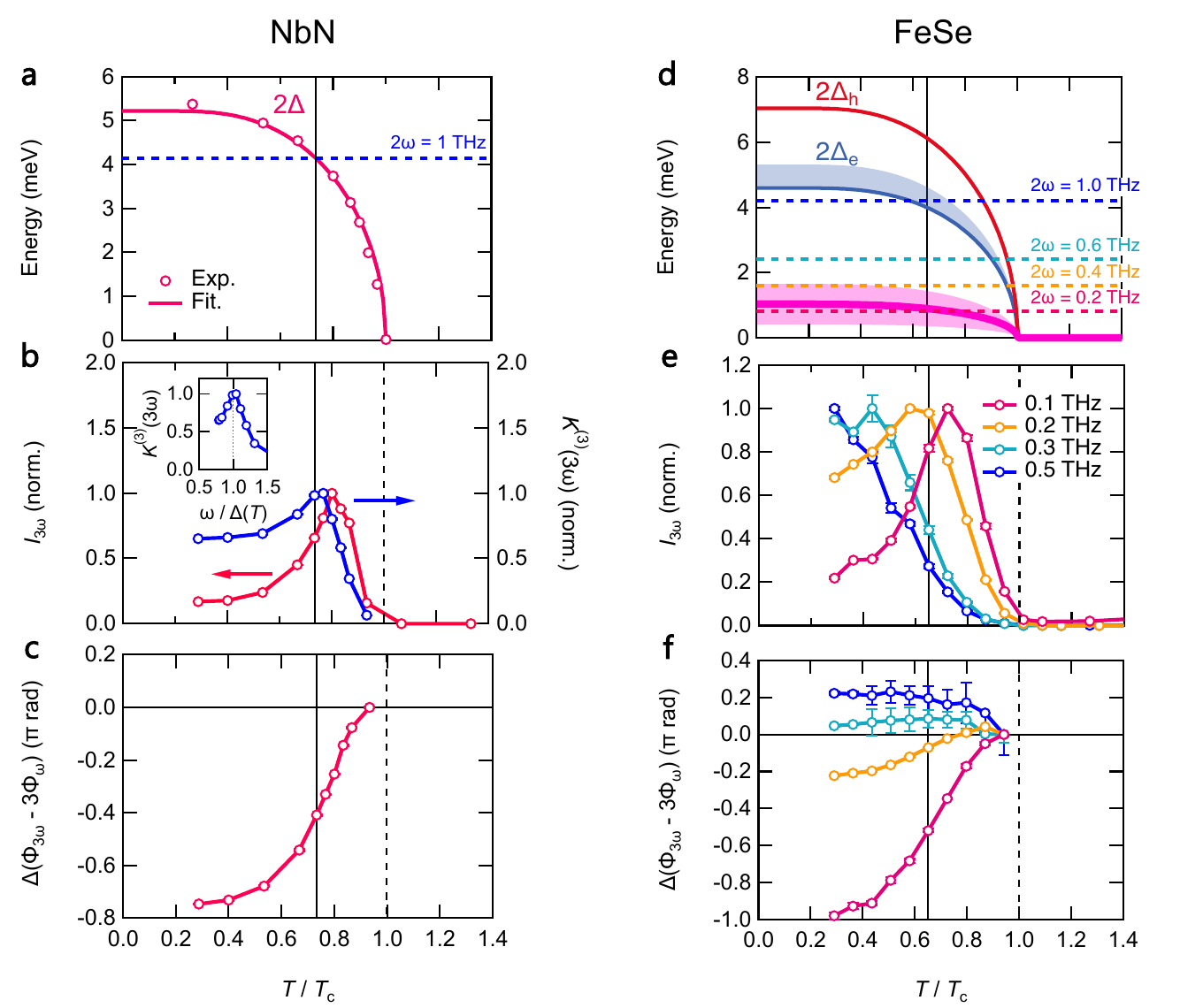}
\caption{{\bf a} Temperature dependence of the superconducting gap energy $2\Delta$ of the NbN sample adopted from ref.\cite{doi:10.1126/science.1254697} (open circles) and fitting curve obtained by a single-band BCS model (solid curve). {\bf b} Temperature dependence of the THG intensity (red) and $K^{(3)}(3\omega) = (I_{3\omega}/I_{\omega}^{3})/\Delta(T)^{4}$ (blue) in NbN normalized by their maximum value for the fundamental freqency of $\omega=0.5\, \text{THz}$. The inset shows $K^{(3)}(3\omega)$ as a function of $\omega/\Delta(T)$, exhibiting a resonant peak at $\omega = \Delta(T)$. {\bf c} Temperature dependence of the relative phase shift of the TH signal with respect to the FH signal in NbN. {\bf d} Estimated temperature dependence of the maximum (antinodal) superconducting gap energy $2\Delta$ of the FeSe sample assuming a single-band BCS-like temperature dependence. The magenta bold curve represents the estimated temperature dependence of the collective mode responsible for the observed THG. {\bf e} Temperature dependence of the TH intensities in FeSe normalized by their maximum value for various incident frequencies (0.1, 0.2, 0.3, or $0.5\, \text{THz}$). {\bf f} Temperature dependence of the relative phase shifts of the TH signal with respect to FH signal in FeSe for each incident frequency.}\label{fig:2}
\end{figure}

For the THG measurements in NbN, we used a narrow-band multicycle THz pulse centered at $0.5\, \text{THz}$ as incident pulse. Figure \ref{fig:2}a shows the temperature dependence of the superconducting gap $2\Delta(T)$ of the NbN sample determined by the complex optical conductivity measurement in the THz frequency range. The solid vertical line represents the temperature at which the two-photon resonance condition of the Higgs mode is satisfied, i.e. $2\omega = 2\Delta(T)$ for the incidence of $\omega = 0.5\, \text{THz}$. As shown by the red open circles in Fig. \ref{fig:2}b, one can identify a resonance peak in the temperature dependence of the TH signal, reproducing the behavior previously reported in ref. \cite{doi:10.1126/science.1254697}. To be precise, in the case of a thin film sample, the effect of the electric field screening inside the sample should be considered to obtain the nonlinear susceptibility of THG as defined by $|\chi^{(3)}|^{2} = I_{3\omega}/I_{\omega}^{3}$ where $I_{\omega}$ denotes the intensity of the transmitted FH component. Furthermore, because $\chi^{(3)}$ contains a proportional factor of $|\Delta(T)|^{2}$ \cite{PhysRevB.92.064508}, we have to evaluate the parameter given by $K^{(3)}(3\omega) = |\chi^{(3)}|^{2}/|\Delta(T)|^{4}$ to accurately identify the resonance energy, as shown by the blue open circles in Fig. \ref{fig:2}b. Without these normalization procedures, the resonance peak in the bare $I_{3\omega} (T)$ plot slightly shifts to the high-temperature side from the true resonance marked by the solid vertical line. The inset of Fig. \ref{fig:2}b represents the same quantity $K^{(3)}(3\omega)$ as a function of $\omega/\Delta(T)$, showing a clear single resonance peak exactly at $\omega = \Delta(T)$ which is accounted for by the paramagnetic coupling between the electromagnetic wave and the Higgs mode \cite{annurev:/content/journals/10.1146/annurev-conmatphys-031119-050813}. 

The presence of resonance also manifests itself in the relative phase shift of the TH with respect to the FH, $\Phi_{3\omega}-3\Phi_{\omega}$, as represented in Fig. \ref{fig:2}c. Here, the phase shift is referenced to the value near $T_{\text{c}}$ ($T = 0.93T_{\text{c}}$) where a sufficient signal-to-noise ratio of the TH signal is obtained. The TH phase shows a negative jump when the temperature is decreased and crosses the Higgs mode resonance ($\omega=\Delta(T)$), which is consistent with the expected behavior of the theory describing the Higgs mode-mediated THG. This result indicates that the temperature dependence of the relative phase shift of the TH signal can be used as an indicator of the collective mode resonance. Importantly, we can identify the resonance by measuring this TH phase shift without performing the above described normalization procedure, making the TH phase measurement effective in capturing the collective mode resonance, in particular when the simultaneous determination of $\Delta(T)$ is difficult, such as in the present case of iron-chalcogenide thin film superconductors.

Now we show in Figs. \ref{fig:2}d-f the temperature dependence of the TH intensity ($I_{3\omega}$) and the relative phase shift in the FeSe sample for four different incident frequencies ($\omega$). The reference temperature is chosen as $0.94T_{\text{c}}$. A phase jump was clearly observed with decreasing temperature for $\omega = 0.1\, \text{THz}$  and $0.2\, \text{THz}$, but not for $\omega \geq 0.3\, \text{THz}$. Concomitantly, the resonance peak in $I_{3\omega}(T)$ is identified only for $\omega = 0.1\, \text{THz}$ and $0.2\, \text{THz}$, although the peak position is shifted to the high temperature side compared to the phase-jump temperature, which is attributed to the competition effect between the internal field screening and the prefactor of $|\Delta(T)|^{4}$ as described for the case of NbN. 

These results indicate that a collective mode exists in the low energy region at least below $2\hbar\omega =2.4\, \text{meV} (\omega = 0.3\, \text{THz})$ at the lowest temperature of $4\, \text{K} ( = 0.29 T_{\text{c}})$. Remarkably, the resonance energy is substantially lower than the superconducting gaps $2\Delta(0)$ of our thin film $\text{FeSe/CaF}_{2}$ sample ($T_{\text{c}}=13.8\, \text{K}$) which are estimated as $2\Delta_{h}(0) = 7.0\, \text{meV}$ for the hole pocket and $2\Delta_{e}(0) = 4.6\, \text{meV}$ for the electron pockets, respectively, by extrapolating the value of the bulk FeSe ($T_{\text{c}}=9\, \text{K}$) determined by STS measurement, $2\Delta_{h}(0) = 4.6\, \text{meV}$ and $2\Delta_{e}(0) = 3.0\, \text{meV}$ \cite{doi:10.1126/science.aal1575}, assuming that $2\Delta_{(h,e)}(0)$ is scaled by $T_{\text{c}}$. For a comparison with the case of NbN, we plot in Fig. \ref{fig:2}d the temperature dependence of the superconducting gaps $2\Delta_{(h,e)}(T)$ with assuming that they are described by the BCS theory as reported in the ARPES measurement \cite{PhysRevX.8.031033}. The possible energy range of the observed low-energy mode is represented by the magenta curve in Fig. \ref{fig:2}d with assuming the BCS-type temperature dependence, which is substantially below $2\Delta_{e}(0)$. 

\section*{ Theoretical analysis}
\begin{figure}
    \centering
    \includegraphics[scale=.45]{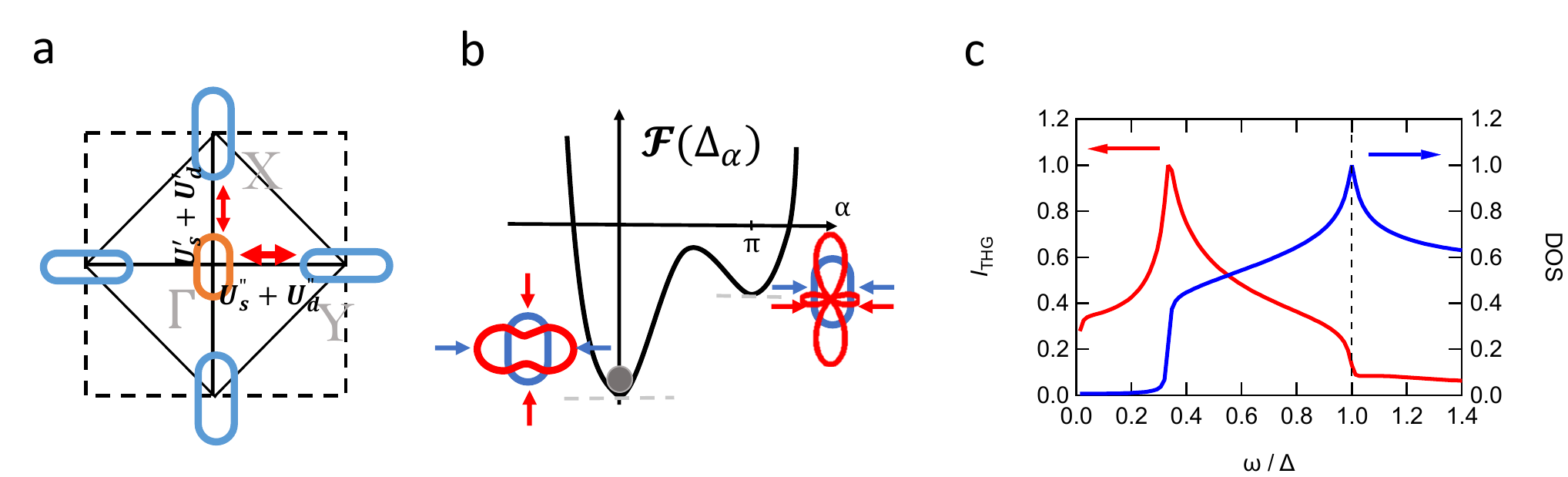}
   \caption{{\bf a} Fermi surfaces at $k_z=0$ of the three-orbital model in the nematic phase. The interactions in the $s$- wave and $d$- wave channels are anisotropic along $\Gamma X$ and $\Gamma Y$. {\bf b} Schematic illustration of the two pairing channels at the hole pocket. The competition between superconductivity and nematic order leads to a free energy with a global minimum corresponding to the ground-state energy and to a local minimum representing the other sub-leading pairing instability.
 Gap structures (red) are shown with respect to the deformed Fermi surface (blue). {\bf c} In red: THG intensity of the Bardasis-Schrieffer mode at $T$=0 corresponding to the subleading state $\Delta_{\mathrm{h}}\propto (-0.5+\cos{2\phi})$. The details of calculations are described in the Supplementary Note 8 and 9. In blue: the density of states corresponding to $\Delta_{\mathrm{h}}\propto(1+0.5\cos{2\phi})$. Both data are normalized by the peak values.}
    \label{fig:theory}
\end{figure}

In this section, we theoretically analyze the possible origin of the observed low-energy mode. In order to phenomenologically reproduce the highly anisotropic superconducting gap structure on both the hole and electron pockets in FeSe, the importance of orbital selective mechanisms has been discussed. Theoretical studies have shown that the largest pairing interaction in the system is between the $\Gamma$-centered hole pocket and the X- and Y-centered electron pockets  \cite{Benfatto_2018,PhysRevLett.120.267001,
doi:10.1126/science.aal1575}, see Fig. \ref{fig:theory}a. This interaction is angle dependent in the band basis and consists of a sign-preserving isotopic s-wave component and a d-wave component \cite{PhysRevLett.120.267001,Benfatto_2018}. Once the nematicity is considered, the interband interaction between the $\Gamma$- and X-pockets becomes grater than that between $\Gamma$- and Y- pockets(Fig.\ref{fig:theory}a). As a consequence, the gaps and pockets become highly anisotropic in the Brillouin zone and behave \textit{as if} the pairing was mediated only by the \textit{d}$_{yz}$ orbital of which the $\Gamma$- and X-pocket are both made of \cite{PhysRevB.98.180503}.

First, we consider the possibility of Higgs modes associated with the hole band or electron bands. As described in the previous section, the observed mode appears substantially below $2\Delta_{h(e)}(0)$, thus it is hardly attributed to the Higgs modes associated with the hole band or electron bands which should show a resonance at $\omega = \Delta(T)$ even for the case of nodal gap structures. 
Second, we examine the effect of quasiparticle excitations, termed charge density fluctuation (CDF). In Raman spectroscopy, this term gives rise to a resonance peak at the gap maxima, as also identified in FeSe\cite{doi:10.1073/pnas.1606562113}, while it can also contribute to THz-THG\cite{PhysRevB.93.180507,PhysRevB.104.174508}. We calculated the TH current for a highly anisotropic OP representing the case of FeSe, where the Higgs mode and CDF contributions appear as an intensity peak and phase jump in the TH signal at the gap maxima, which contradicts the experimental result (see Supplementary Note 7 for more details). 

It should be noted here that if there exists a small but finite gap around the nodal region as observed in STS measurements\cite{doi:10.1126/science.aal1575}, the Higgs mode or CDF may give rise to a resonant structure in the TH signal at the gap minima in addition to the gap maxima. We examined this possibility by taking into account the anisotropic gap structure. As described in Supplementary Note 7, the calculated spectra indeed exhibit a resonant structure at the proximity of gap minima (Fig.S10). This result suggests possible contributions of the Higgs mode and CDF to the observed low-energy mode, but this picture still fails to explain the absence of high-energy resonance at the gap maxima. Accordingly, we are led to consider other mechanisms as the origin of the low-energy mode that overwhelms the Higgs mode and CDF contributions.  

We then consider the possibility of the BS mode associated with the hole pocket as described below, following the picture inspired by Refs.\cite{PhysRevLett.120.267001,Huang_2018}(see Supplementary Note 8 for more details).
We parametrize the hole gap as made up of two components $\{\Delta_1,\Delta_2 \cos(2\phi)\}$ corresponding to the $s$-wave and $d$-wave pairing channels, respectively. By considering the primary interactions in the system, we write an effective linearized self-consistent gap equation for the two components as in ref.\cite{Huang_2018} in a matrix form as:
\begin{equation}
\binom{\Delta_1}{\Delta_2}=-N_0 \left(\begin{array}{cc}
V_s & V_n \\
V_n & V_d
\end{array}\right)\binom{\Delta_1}{\Delta_2},
\end{equation}

where $V_s$ and $V_d$ are the combination of the isotropic and anisotropic parts of the interactions $V^{he_X}$ and $V^{he_Y}$ between the hole and electronic pockets. The off diagonal terms, i.e. the mixing between the $s$- and $d$- wave representations arises due to the nematic distortion of the hole and electron pockets. In the above coupled-BCS gap equation, $N_0$ is independent of the gap. 

We then solve the gap equation by diagonalizing $\hat{V}$ and obtain the effective pairing interaction in the two eigenchannels: one is an $s$-wave dominated channel and sign preserving, corresponding to the form factor of $f_0(k)=1+r \cos{2\phi}$ which represents a gap distorted such that the minima of the gap are along the direction of greater distortion of the pocket. This represents the ground state of our system. The second corresponds to a $d$-wave dominated one with $f_1(k)=\cos{2\phi}-r'$. In the latter channel, the sign change turns in a gap that whose maxima are oriented along the direction of grater distortion of the pocket. This represents a sub-leading pairing channel in our description, although it is an admixture of the $s$-wave and $d$-wave pairing channels, as schematically drawn in Fig. \ref{fig:theory}b. The picture discussed here is in agreement with ref.\cite{Mishra_2016}. In such a case, a collective excitation corresponding to the BS mode, i.e. the relative phase oscillation between the $s$-wave-like ground state and the $d$-wave-like subleading pairing state, is anticipated to exist.
We calculated the TH current mediated by this BS mode under the driving field of frequency $\omega$, and plotted the intensity of TH with changing $\omega$ in Fig. \ref{fig:theory}c (see the Supplementary Note 9 for more details). Also shown is a density of state for the gap function for the hole band. A sharp resonance mediated by the ${\it hybridized}$ BS mode appears slightly below the edge of the gap minima, lacking a resonance structure at the gap maxima, which reasonably well reproduces the experimental observation.   \\

\section*{Polarisation dependence of the THG}\label{subsec:PolDep}
To further investigate the character of the observed low-energy mode, we measured its polarisation dependence. The details of the measurement scheme are described in Supplementary Note 6.
As shown in Fig. \ref{fig:4}a, the TH signal is almost isotropic with respect to the crystal axis. Figure \ref{fig:4}b represents the power spectrum of the transmitted waveforms for the incident frequency of 0.3 THz. We can see that the TH signal appears only in the polarisation direction parallel to the incident wave and that the perpendicular component is absent within our detection sensitivity. 

We confirmed the crystallinity and the uniformity of our epitaxially grown FeSe sample at room temperature by XRD measurements (see Supplementary Note 2), from which we rule out the possibility that the observed isotropic behavior of the TH signal arises from the inhomogeneity of the crystal axis orientation. However, it is recognized that FeSe samples have a twinned structure consisting of nematic domains below the structural transition temperature, rotated by 90 degrees relative to each other, even in the superconducting state. We could not resolve those domain structures in our sample by an optical microscope, indicating that the typical domain size is smaller than the visible wavelength and thus more than 3-orders of magnitude smaller than the THz probe wavelength. Therefore, the observed isotropy could be attributed to the effect of averaging over the multiple domains of the sample. 

To compare, we calculated the polarisation dependence of the TH nonlinear current originating from the BS mode as shown in Fig. \ref{fig:4}c. The calculation shows that the mode exhibits $C_2$ symmetry with respect to the crystal axis, while exhibiting a substantially smaller perpendicular component (the details of the calculation are described in Supplementary Note 9). Taking into account the twinned multidomain structure of the sample, we then averaged the current over domains, the results of which are represented in Fig. \ref{fig:4}d. Although there still remains a finite anisotropy in the polarisation dependence, one can see that the overall anisotropy is largely suppressed. More importantly, the perpendicular component is substantially smaller than the parallel component, in accordance with the experimental observations. The remaining anisotropy might be further smeared out by the effect of impurities or defects.

\section*{Discussion}\label{sec:summary}
As discussed in refs.\cite{fernandesNematicityProbeSuperconducting2013,PhysRevLett.108.247003}, in the absence of strong nematicity that linearly mixes the $s$-wave and $d$-wave states, the system can stabilize in a time-reversal symmetry-breaking (TRSB) $s+e^{i\alpha}d$ state with a relative phase $\alpha$ different from $0$ or $\pi$ at low temperatures. This SC-to-SC transition should be signaled by the softening of the collective phase fluctuation between the X and Y bands, correspnding to the Leggett mode, as suggested in Refs.\cite{ PhysRevB.98.064508}. 

Looking at the temperature dependence of the $I_{3\omega}$ shown in Fig. \ref{fig:2}e, it seems that the system is consistently in the same phase within the superconducting regime for all measured temperature, as there is no indication of mode softening with lowering the temperature. It is also worth noting that recent $\mu$SR experiment in bulk FeSe indicated the development of TRSB state just below $T_{\rm c}$ \cite{doi:10.1073/pnas.2208276120}.
Even for such a case, we expect the coupling of the Leggett mode with the light to be small in the present case, because the Leggett mode between the two electron pockets proposed in Refs.\cite{Huang_2018, PhysRevB.98.064508} would couple with light as $( c^{e_X}_0-c^{e_Y}_0)^2$, where $c^{e_X}_0$ and $c^{e_Y}_0$ are the coupling between light and each band (see ref.\cite{PhysRevB.95.104503}), being small when the two pockets are of the same character. Therefore, it is unlikely to attribute the observed mode to the softened Leggett mode associated with TRSB. 

In the case of FeSe, the nematic order is not weak and the bands are strongly anisotropic, which would primarily lead to the $s$+$d$ state.
Our results corroborate such a mixed symmetry state through the observation of the low-energy mode substantially below the superconducting gap,which is ascribed to the BS mode.
The BS mode between the $s+d$-wave ground state and the subleading pairing channel can also be interpreted as an intra-pocket phase fluctuation between the $s$- and $d$- wave component.
In this sense, one may also call this mode as the {\it Leggett} mode in a single band system, as noted in Ref.\cite{Huang_2018} and by Leggett himself \cite{leggettNumberPhaseFluctuationsTwoBand1966}.
As demonstrated in this work, THz nonlinear spectroscopy enables a unique access to the collective modes, providing a deep insight into the symmetry and the structure of superconducting order parameter, and paves a new pathway to investigate further exotic superconducting states including TRSB and topological superconductors.

\section*{Methods}\label{sec:methods}
\subsection*{Sample growth and characterization}\label{subsec:method_sample}
The thin film FeSe samples oriented along the c-axis were epitaxially grown on a $\text{CaF}_{2}$ substrate by pulsed laser deposition method using KrF laser \cite{Imai_2010,doi:10.1073/pnas.1418994112}. The samples were characterized by transport measurements and X-ray diffraction measurements (see Supplementary Note 1 and 2 for the details). The THz-THG measurements were carried out on a $65\, \text{nm}$-thick FeSe film grown on a $0.5\, \text{mm}$-thick $\text{CaF}_{2}$ substrate, whose $T_{\text{c}}$ was $13.8\, \text{K}$. The polarisation dependence of the THG signal was measured with a sample grown in the same conditions. These films are compressively strained due to the substrates, according to which $T_{\text{c}}$ is higher than that of bulk FeSe samples \cite{10.1063/1.4826945}. The results of the optical characterization of the sample are shown in Supplementary Note 3 and 5.

\vskip\baselineskip

The thin film of NbN sample ($24\, \text{nm}$ in thickness) used in the THG measurements as a reference was fabricated on a $0.5\, \text{mm}$-thick MgO substrate whose $T_{\text{c}}$ was $15\, \text{K}$. The temperature dependence of the superconducting gap of this sample is adopted from ref.\cite{doi:10.1126/science.1254697}.

\subsection*{THG experiments}\label{subsec:method_experiments}
For THG measurements, we first generated an intense broad-band THz pulse via a tilted pulse-front scheme utilizing a $\text{LiNbO}_{3}$ crystal \cite{Watanabe:11}, driven by a regenerative amplified Ti:sapphire laser (pulse duration of $100\, \text{fs}$, central photon energy of $\hbar\omega = 1.55\, \text{eV}$, and repetition rate of $1\, \text{kHz}$). Then, a narrow-band multicycle THz pulse was obtained using a bandpass filter (BPF). Four BPFs were used to generate the different center frequencies at 0.1, 0.2, 0.3, or $0.5\, \text{THz}$, respectively. To extract the TH signal with suppressing the residual FH ($\omega$) component, $3\omega$-BPFs were inserted after the sample. The time-domain waveform of THz pulse was detected by electro-optic (EO) sampling in a ZnTe (110) crystal with a thickness of $2\, \text{mm}$. The sample was mounted on a copper sample holder with a tapered hole ($6\, \text{mm}$ in diameter). Temperature-dependence measurements were performed using a helium-flow cryostat.

To measure the dependence of THG on the electric field polarisation, several wire-grid polarizers (WGPs) were used. For the details of the polarisation-resolved THG measurements, see Supplementary Note 6.

\newpage
\section*{Figures}\label{sec6}

\begin{figure}[h]
\centering
\includegraphics[width=0.9\textwidth]{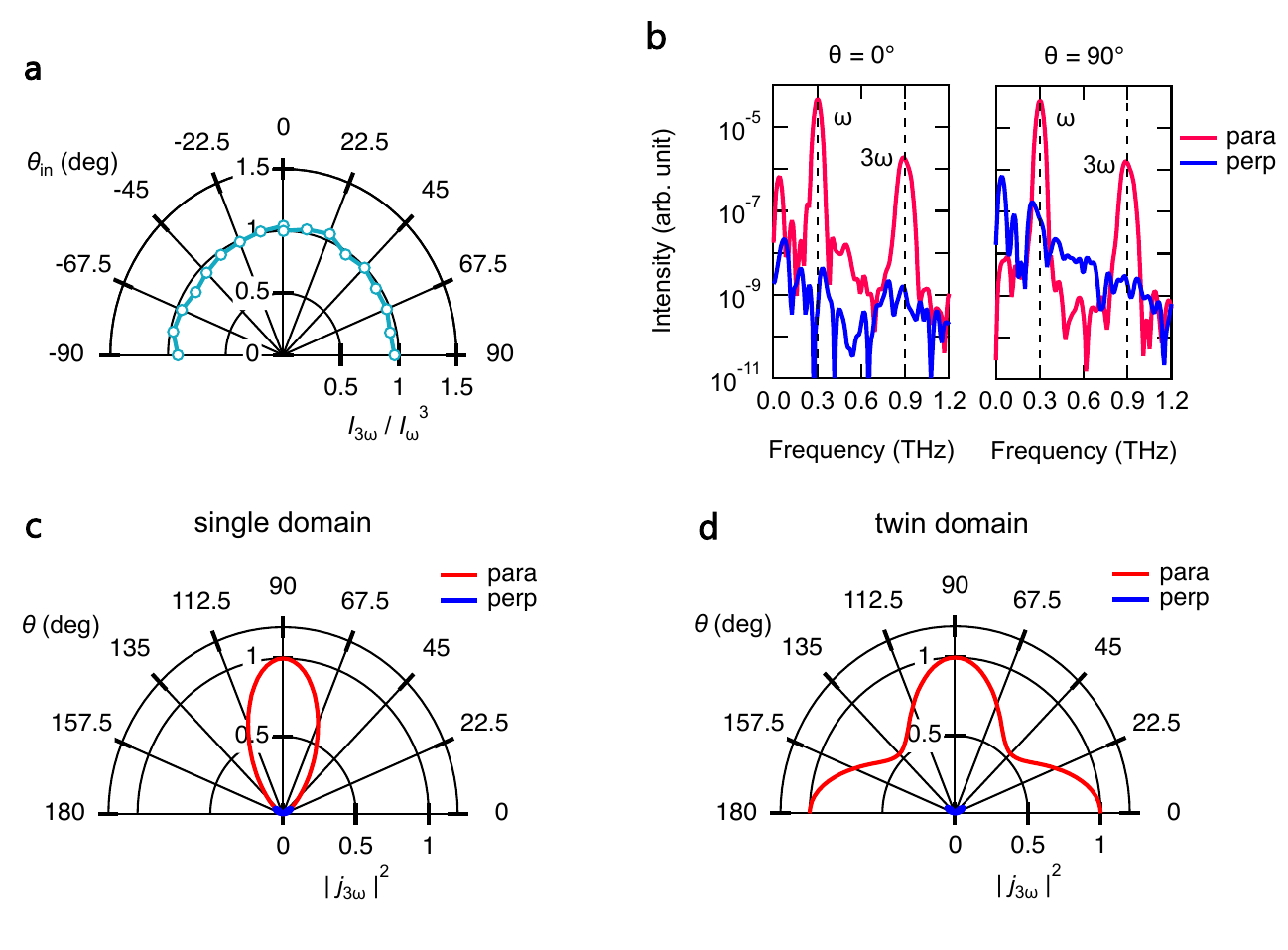}
\caption{{\bf a} Incident polarisation angle dependence of the TH signal  normalized by the fundamental component, $I_{3\omega}/I_{\omega}^{3}$ for $\omega = 0.3\, \text{THz}$ incidence. The data is normalized by the maximum value. {\bf b} Power spectrum for the transmitted THz pulse for $\omega = 0.3\, \text{THz}$ incidence. The parallel ($E^{\parallel}$) and the perpendicular ($E^{\perp}$) components with respect to the incident THz polarisation are shown.{\bf c} Calculated polarisation dependence for the Bardasis-Schrieffer mode. {\bf d} Calculated polarisation dependence for the Bardasis-Schrieffer mode averaged over twinned multiple domains. }\label{fig:4}
\end{figure}

\clearpage

\section*{Acknowledgements}
This paper was in part support by JST CREST grant No. JPMJCR19T3, Japan and the JSPS KAKENHI (No. JP25H01251). R.S. and D. M. acknowledge the support by the Max Planck-UBC-UTokyo Center for Quantum Materials.

\section*{Data availability}
The data that support the findings of this study are available from the corresponding authors upon reasonable request.

\section*{Contributions}
 R.S. conceived the project of this study. H. M. carried out the optical experiments. S. N, H. M. and D. M. conducted the theoretical modeling with feedback from R.S. T.K. and A.M. fabricated and characterized the sample. H. M., S. N, and R.S. analyzed the data and wrote the manuscript with input from all the coauthors.

\section*{Corresponding authors}
Correspondence to Ryo Shimano (shimano@phys.s.u-tokyo.ac.jp)


\pagebreak
\setcounter{equation}{0}
\setcounter{figure}{0}
\setcounter{table}{0}
\setcounter{page}{1}
\setcounter{enumiv}{0}
\makeatletter

\renewcommand{\thepage}{S\arabic{page}}
\renewcommand{\thesection}{S\arabic{section}}
\renewcommand{\thetable}{S\arabic{table}}
\renewcommand{\thefigure}{S\arabic{figure}}
\renewcommand{\theequation}{S\arabic{equation}}
\renewcommand{\bibnumfmt}[1]{[S#1]}
\renewcommand{\citenumfont}[1]{S#1}

\begin{center}
{\LARGE \phantom{a} \\ Supplementary Information \\ \phantom{a} \\ \bf{A new collective mode in an iron-based superconductor with electronic nematicity} \\ \phantom{a} \\ \phantom{a} \\ \phantom{a}}
\end{center}

\section*{Supplementary Note 1. Temperature dependence of the DC resistance}\label{sec:SI_RT}

\begin{figure}[h]
\centering
\includegraphics[width=0.8\textwidth]{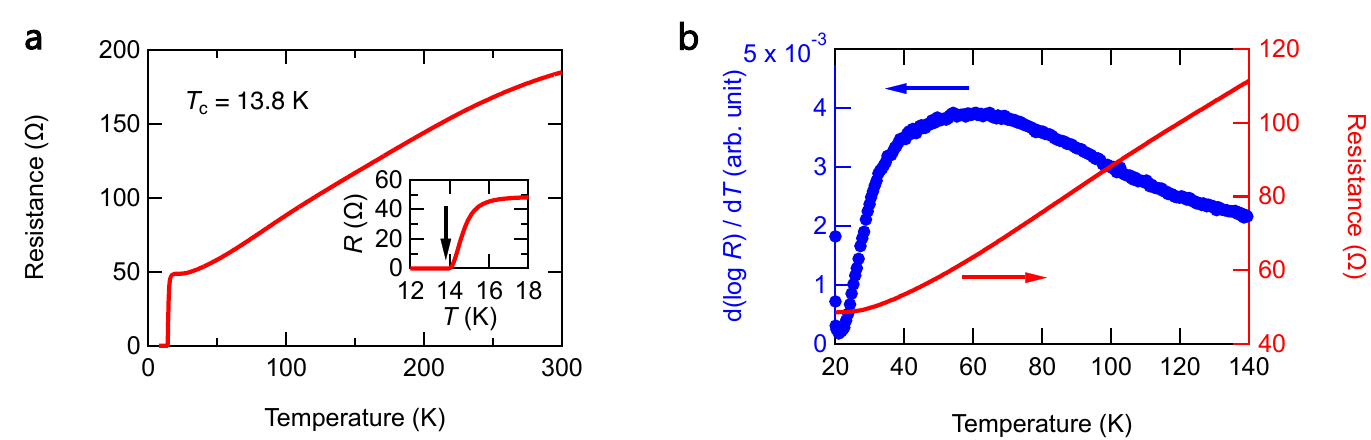}
\caption{{\bf a} Temperature dependence of the DC resistance of a thin film of FeSe sample. The black arrow in the inset shows the superconducting critical temperature defined as the onset of zero resistivity. {\bf b} Temperature dependence of resistance $R$ (red curve) and $d (\log{R})/dT$ (blue circles).}\label{fig:S1}
\end{figure}

The temperature dependence of the DC resistance of our thin film sample is presented in Fig. \ref{fig:S1}(a). The superconducting critical temperature is evaluated as $T_{\text{c}}=13.8\, \text{K}$ from the onset of zero-resistivity temperature. The tetragonal-orthorhombic structural transition temperature $T_{\text{s}}$, which is associated with the Fe $3d_{xz}$ / $3d_{yz}$ orbital ordering \citesi{SI_PhysRevB.90.121111}, is evaluated from the temperature derivatives of the logarithms of resistance as shown in Fig. \ref{fig:S1}(b) following the procedure adopted in ref.\citesi{SI_Imai2017}. $T_{\text{s}}$ is then assigned as a broad peak appearing at $T \approx 60\, \text{K}$.

\section*{Supplementary Note 2. X-ray diffraction analysis}\label{sec:SI_XRD}

\begin{figure}[h]
\centering
\includegraphics[width=0.9\textwidth]{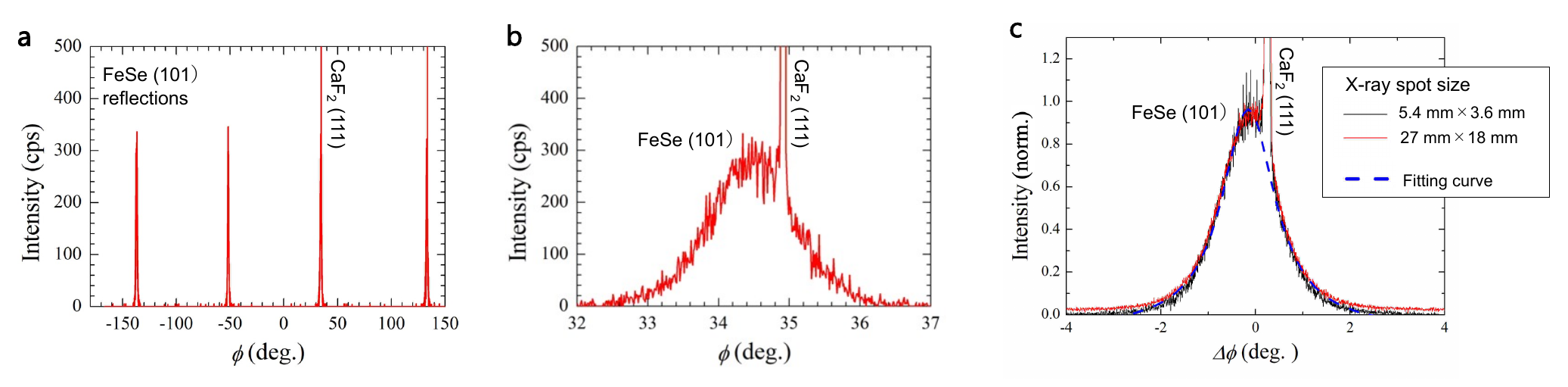}
\caption{{\bf a} XRD patterns of the $\phi$-scan of the (101) reflection from the FeSe/$\text{CaF}_{2}$ sample. {\bf b} Enlarged view of ({\bf a}) around the peak at $\phi = 34.5\degree$. {\bf c} The FeSe (101) peak in the $\phi$-scan (red and black curves) taken with different X-ray spot sizes and the fitted result (dashed curve). The $\phi$-scans are normalized by the FeSe (101) peaks.}\label{fig:S2}
\end{figure}

The X-ray diffraction pattern of the epitaxial thin film FeSe/$\text{CaF}_{2}$ sample is measured at room temperature to confirm the crystal axes orientations. The sample used in this XRD characterization is obtained under the same growth conditions with the sample used for the terahertz (THz) measurements in the main text. The obtained $\phi$-scans of the (101)- reflection are shown in Figs. S2(a-c). As shown in Fig. \ref{fig:S2}(a), a clear four-fold symmetry reflection was observed. Figure \ref{fig:S2}(b) is an enlarged plot of Fig. \ref{fig:S2}(a) around $\phi = 34.5\degree$. The full width at half maximum of the peak is $1.37\degree$, indicating the high crystallinity of the sample. Further detailed characterizations of thin film samples fabricated by the same method can be found in ref.\citesi{SI_10.1063/1.4826945}. To examine the uniformity of the sample in a larger scale, we measured $\phi$-scans of the FeSe $(101)$ peak with different X-ray spot sizes as shown in Fig. \ref{fig:S2}(c). The totally same profile of the $\phi$-scans for different spot sizes ensures the uniformity of the sample in this spatial scale.

\section*{Supplementary Note 3. Optical conductivity}\label{sec:SI_OpticalConductivity}

We have characterized the optical properties of our thin film FeSe/$\text{CaF}_{2}$ the sample using THz time-domain spectroscopy and obtained the complex optical conductivity without resorting to Kramers-Kronig analysis. The broadband THz probe pulse was generated from a p-type InAs (111) crystal illuminated by a NIR femtosecond pulse from the regenerative amplified Ti:sapphire laser (pulse duration of $100\, \text{fs}$, central photon energy of $\hbar\omega = 1.55\, \text{eV}$, repetition rate of $1\, \text{kHz}$). The waveform of the transmitted probe pulse after the sample was measured by electro-optic (EO) sampling using a ZnTe (110) crystal with a thickness of $2\, \text{mm}$ the same as that used in the THG measurements.

\begin{figure}[h]
\centering
\includegraphics[width=0.8\textwidth]{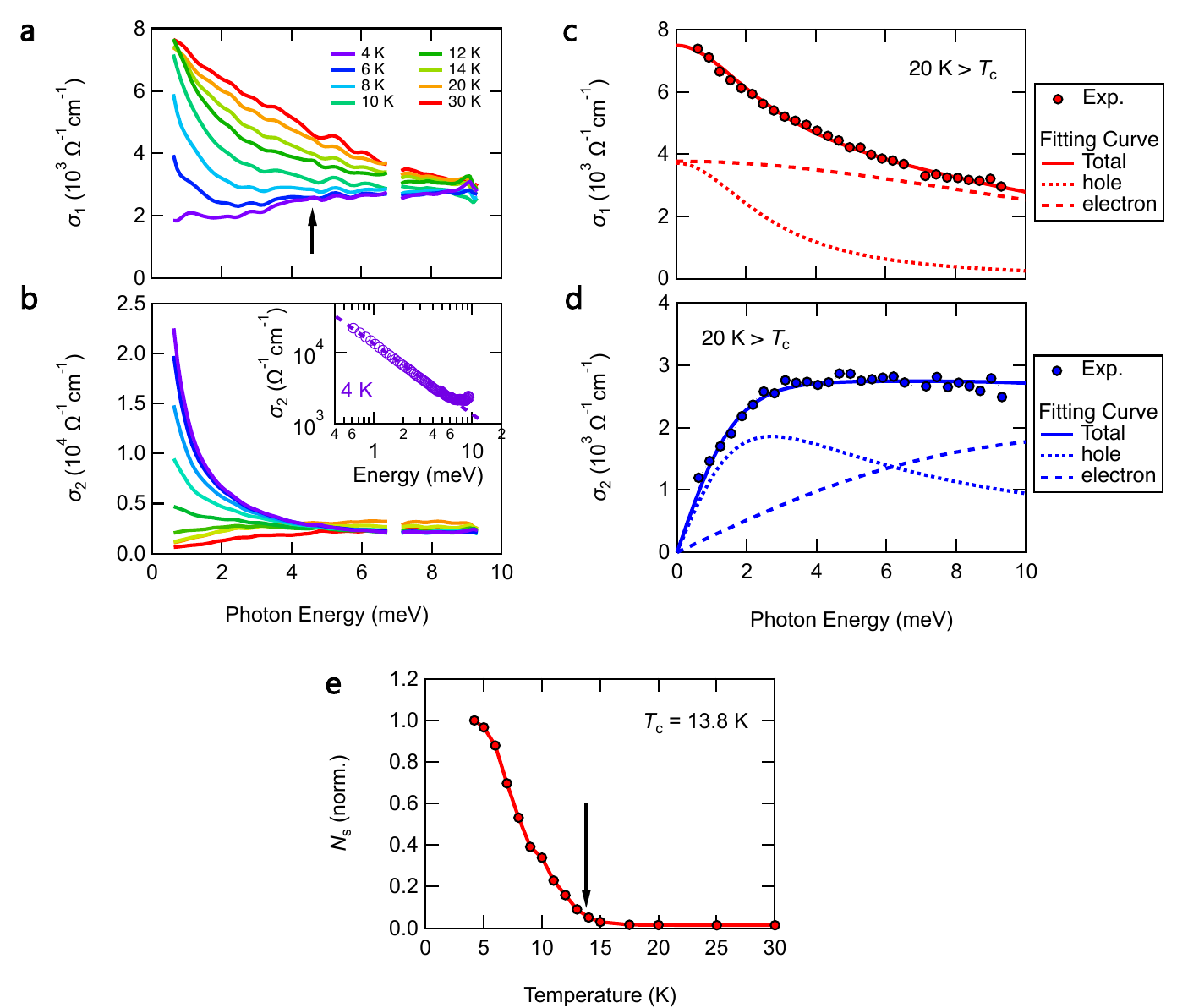}
\caption{{\bf a, b} Real ($\sigma_{1}$) and imaginary ($\sigma_{2}$) part of the optical conductivity of our FeSe sample. The black arrow in ({\bf a}) indicates an energy where a downturn appears at $4\, \text{K}$. The inset of ({\bf b}) shows a logarithmic plot of $\sigma_{2}$ at $4\, \text{K}$ (open circles). The dashed line represents the $1/\omega$ function. {\bf c, d} Real and imaginary part of the optical conductivity at $20\, \text{K} (> T_{\text{c}})$ (circles). The fitted results with two-component Drude model are also shown: the dotted lines indicate the hole carrier contributions, and the dashed lines indicate the electron carrier contributions, and solid lines indicate the total fitting curves. {\bf e} Temperature dependence of the superfluid density obtained by using the curve fitting of the $\sigma_{2} (\omega)$ with the two-fluid model. The arrow indicates $T_{\text{c}}$.}\label{fig:S3}
\end{figure}

The measured optical conductivity of the FeSe thin film sample is shown in Figs. \ref{fig:S3}(a) and (b). Below $T_{\text{c}} = 13.8\, \text{K}$, the missing spectral weight in the real part of the optical conductivity, $\sigma_{1}$, was observed up to $\sim 8\, \text{meV}$, indicating the opening of the superconducting gap with an approximate gap size of $\sim 8\, \text{meV}$. In addition, a downturn was observed below $\sim 4.5 \text{meV}$ at $4\, \text{K}$, suggesting the presence of a smaller superconducting gap. These values are consistent with the superconducting gap in our FeSe sample, $2\Delta_{\mathrm{h}} = 7.1\, \text{meV}$ and $2\Delta_{\mathrm{e}} = 4.6\, \text{meV}$, respectively, estimated from those measured by STS in previous research \citesi{SI_doi:10.1126/science.aal1575}, assuming the scaling relation $\Delta \propto T_{\text{c}}$. The residual spectral weight in the low energy region below $2\Delta_{\mathrm{e}}$ at the lowest temperature limit ($4\, \text{K}$ in our setup) is plausibly attributed to the remaining quasiparticle weight reflecting the nodal-like structure of the gap function \citesi{SI_PhysRevB.84.134522}.
The optical conductivity at $20\, \text{K}$, in the normal state above $T_{\text{c}}$, is represented in Figs. \ref{fig:S3}(c) and (d). According to the previous study using THz magneto-optical Faraday rotation spectroscopy \citesi{SI_PhysRevB.100.035110}, the charge carrier response in FeSe is well described by the two-component Drude model up to $\sim 100\, \text{K}$, whose optical conductivity is written as

\begin{equation}
    \label{eq:2Drude_OC}
    \sigma(\omega) = \frac{\epsilon_{0}\omega_{\text{p},1}}{\gamma_{1}-i\omega} + \frac{\epsilon_{0}\omega_{\text{p},2}}{\gamma_{2}-i\omega}
\end{equation}

where $\omega_{\text{p},i}$ and $\gamma_{i}\ (i=1,2)$ represent the plasma frequency and the scattering rate of the $i$-th component, respectively, and $\epsilon_{0}$ denotes the dielectric constant. 
From the above two-component Drude fit, we obtained the parameters of our sample at $20\, \text{K}$ as follows.

\begin{align*}
    \omega_{\text{p,h}} &= 274\, \text{meV} \pm 6\, \text{meV}, &
    \gamma_{\text{h}} &= 2.72\, \text{meV} \pm 0.07\, \text{meV}, &&& \text{(hole)} \\
    \omega_{\text{p,e}} &= 634\, \text{meV} \pm 3\, \text{meV}, &
    \gamma_{\text{e}} &= 14.3\, \text{meV} \pm 0.3\, \text{meV}. &&& \text{(electron)} \\
\end{align*}

The logarithmic plot of the imaginary part of the optical conductivity ($\sigma_{2} (\omega)$) below $T_{\text{c}}$ is shown in the inset of Fig. \ref{fig:S3}(b), showing the behavior of $\sigma_{2} (\omega) \propto N_{s}/\omega$  up to $\omega \sim 6\, \text{meV}$, with $N_{s}$ representing the superfluid density. The temperature dependence of the superfluid density is obtained using a two-fluid model fit to $\sigma_{2}(\omega)$ in the frequency range from $0.06\, \text{meV}$ to $4\, \text{meV}$, with normal carrier contribution assumed to take the same two-Drude model described above. The result is shown in Fig. \ref{fig:S3}(e).

\section*{Supplementary Note 4. Effects of multiple reflection inside the substrate}\label{sec:SI_MultiReflection}

\begin{figure}[h]
\centering
\includegraphics[width=0.5\textwidth]{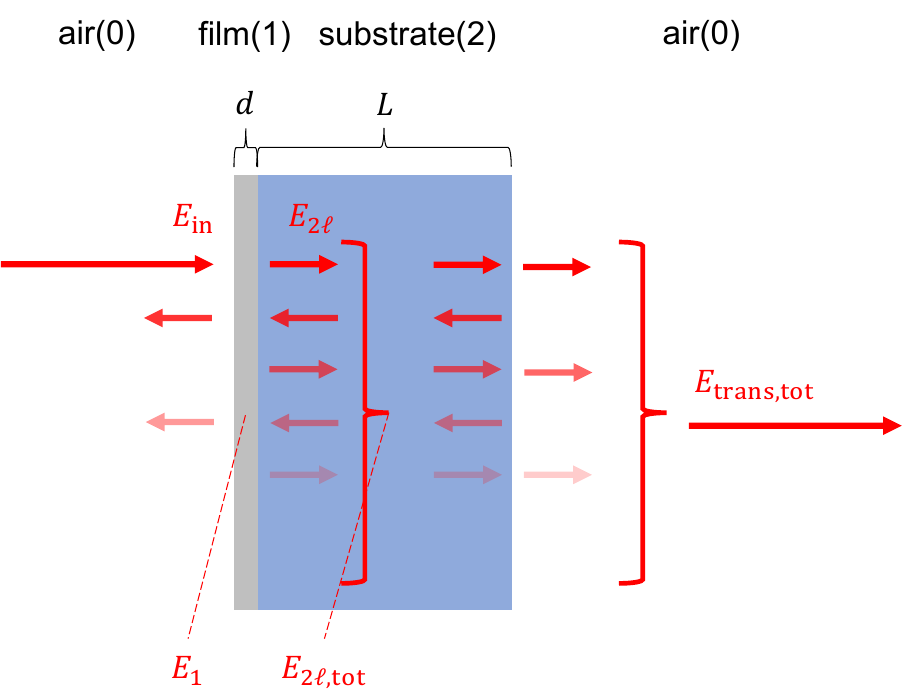}
\caption{Schematic picture of the multiple reflection in the substrate.}\label{fig:S4}
\end{figure}

Since the sample we used is a thin film of FeSe grown on a submillimeter thickness substrate, the measured waveform of the transmitted THz pulse is affected by the multiple reflection in the substrate as illustrated in Fig. \ref{fig:S4}. Note that the THz pulse was incident from the film side. If the pulse width is much shorter than one round trip time for the light in the substrate, the effect of multiple reflection in the substrate can be eliminated by limiting the measurement time window to be shorter than the round trip time. However, we could not use this method because the duration of the multicycle THz pulse used in the THG experiments does not meet the above condition. Consequently, in this section, we examine the effect of multiple reflections on our THG measurements.

Here we adopt the thin film approximation, in which we assume a uniform electric field inside the film with the relationship $E_{1}=E_{2\ell,\text{tot}}$, where $E_{1}$ represents the electric field inside the film, and $E_{2\ell,\text{tot}}$ represents the sum of the electric fields of the transmitted pulse just exiting the film/substrate interface (see Fig. \ref{fig:S4}). Using this thin-film approximation, the reflection coefficient $r$ at the film/substrate boundary incident from the substrate side is given as

\begin{equation}
    \label{eq:r_S/F/A}
    r = -\frac{1-n_{s}+Z_{0}d\sigma}{1+n_{s}+Z_{0}d\sigma}
\end{equation}

where $n_{s}$ is the refractive index of the substrate, $d$ is the thickness of the film, $\sigma$ is the optical conductivity of the film and $Z_{0} = \sqrt{\mu_{0}/\epsilon_{0}}$ is the vacuum impedance. For details of the thin-film approximation, see Supplementary Note 5.

The relationship between $E_{2\ell}$, the electric field of the transmitted pulse that just exits the film/substrate interface before experiencing multiple reflections in the substrate, and $E_{\text{trans,tot}}$, the total electric field of the transmitted pulse exiting the substrate, is described by the Fresnel equation:

\begin{equation}
    \label{eq:Etrans_tot}
    E_{\text{trans,tot}} = E_{2\ell}\frac{t_{2}t_{20}}{1-r_{20}t_{2}^{2} r}
\end{equation}

where $t_{20}=2n_{s}/(n_{s}+1)$ describes the transmission across the substrate/air interface from the substrate side, $r_{20} = (n_{s}-1)/(n_{s}+1)$ describes the reflection at the substrate/air interface from the substrate side, $t_{2} = \exp{(i\omega L n_{s}/c)}$ describes the phase factor for propagation through the substrate of thickness L. Next, the relationship between $E_{2\ell}$ and $E_{2\ell,\text{tot}}$, the total electric field of the transmitted pulse just exiting the film/substrate interface considering the multiple reflection in the substrate, is described by the following Fresnel equation.

\begin{equation}
    \label{eq:E2l_tot}
    E_{2\ell,\text{tot}} = E_{2\ell}\frac{1}{1-r_{20}t_{2}^{2} (1+r)}
\end{equation}

Using these equations and with the assumption of the thin film approximation, $E_{1}=E_{2\ell,\text{tot}}$, the relationship between $E_{1}$ and $E_{\text{trans,tot}}$ is given as follows:

\begin{equation}
    \label{eq:E1_tot}
    E_{1,\text{tot}} 
    = E_{\text{trans,tot}}\left(t_{2}t_{20}\frac{1-r_{20}t_{2}^{2} (1+r)}{1-r_{20}t_{2}^{2} r} \right)
    = E_{\text{trans,tot}} A,
\end{equation}

where the coefficient $A = t_{2} t_{20} (1-r_{20} t_{2}^{2} (1+r))/(1-r_{20} t_{2}^{2} r)$ denotes the ratio between the electric field inside the film and the electric field of the transmitted pulse measured. We note that the relationships derived so far are valid for both fundamental and TH frequency components as long as we assume the thin-film approximation. Although the incident pulse does not exist for the TH component, the relationship $E_{1}=E_{2\ell,\text{tot}}$ holds as long as we assume the thin-film approximation and the same derivation is applicable for the relationship between $E_{1}$ and $E_{\text{trans,tot}}$.

\begin{figure}[h]
\centering
\includegraphics[width=0.7\textwidth]{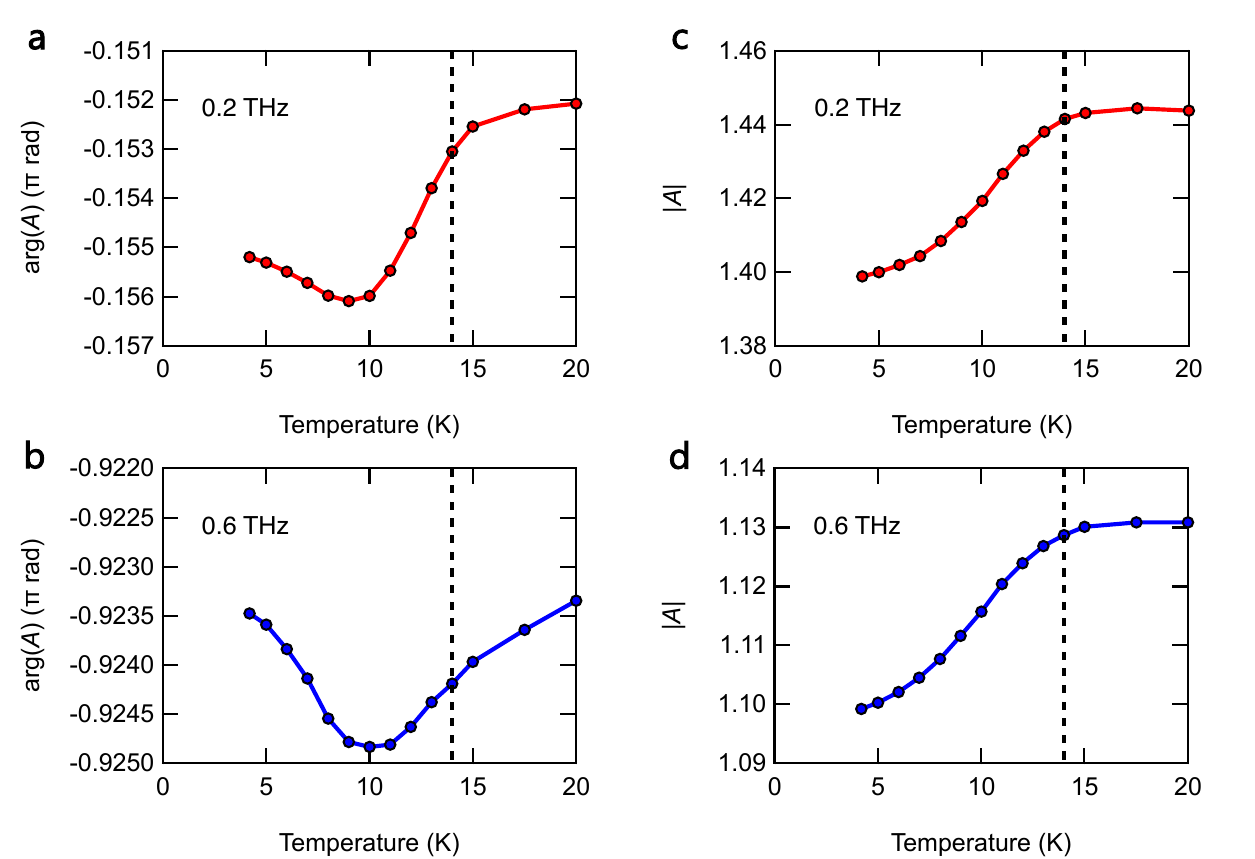}
\caption{The temperature dependence of {\bf a, b} the phases and {\bf c, d} the amplitudes of the ratio $A$ between the electric field inside the film and the electric field of the transmitted pulse just to the right of the film/substrate interface at {\bf a,c} $0.2\, \text{THz}$  and {\bf b,c} $0.6\, \text{THz}$  calculated from the optical conductivity.}\label{fig:S5}
\end{figure}

Since the coefficient $A$ depends on the temperature dependent conductivity of the film, the temperature dependences of the intensity and phase of the measured transmitted pulse are not equal to those inside the film. Thus, we evaluated the temperature dependence of this coefficient at frequencies from $0.15\, \text{THz}$ to $2\, \text{THz}$ using the temperature dependence of the optical conductivity of the films measured at these frequencies. The results at 0.2 and $0.6\, \text{THz}$ for the FeSe sample are shown in Figs. \ref{fig:S5}(a-d). As shown in Figs. \ref{fig:S5}(a, b), the phases of $A$ change less than $5\times 10^{-3}$  $\pi\, \text{rad}$ below $20\, \text{K}$ at both 0.2 and $0.6\, \text{THz}$, which are negligible for our THG measurements. The amplitude change of $A$ is less than 4\% below $20\, \text{K}$ for both 0.2 and $0.6\, \text{THz}$, as shown in Figs. \ref{fig:S5}(c, d), which is also substantially small. Accordingly, with a good enough accuracy, we can approximate the temperature dependence of the transmitted electric field $E_{\text{trans,tot}}$, which is the quantity measured in the THG experiments, to that of the electric field inside the thin film sample, $E_{1}$.

\section*{Supplementary Note 5. Thin film approximation}\label{sec:SI_ThinFilmApprox}

\begin{figure}[h]
\centering
\includegraphics[width=0.5\textwidth]{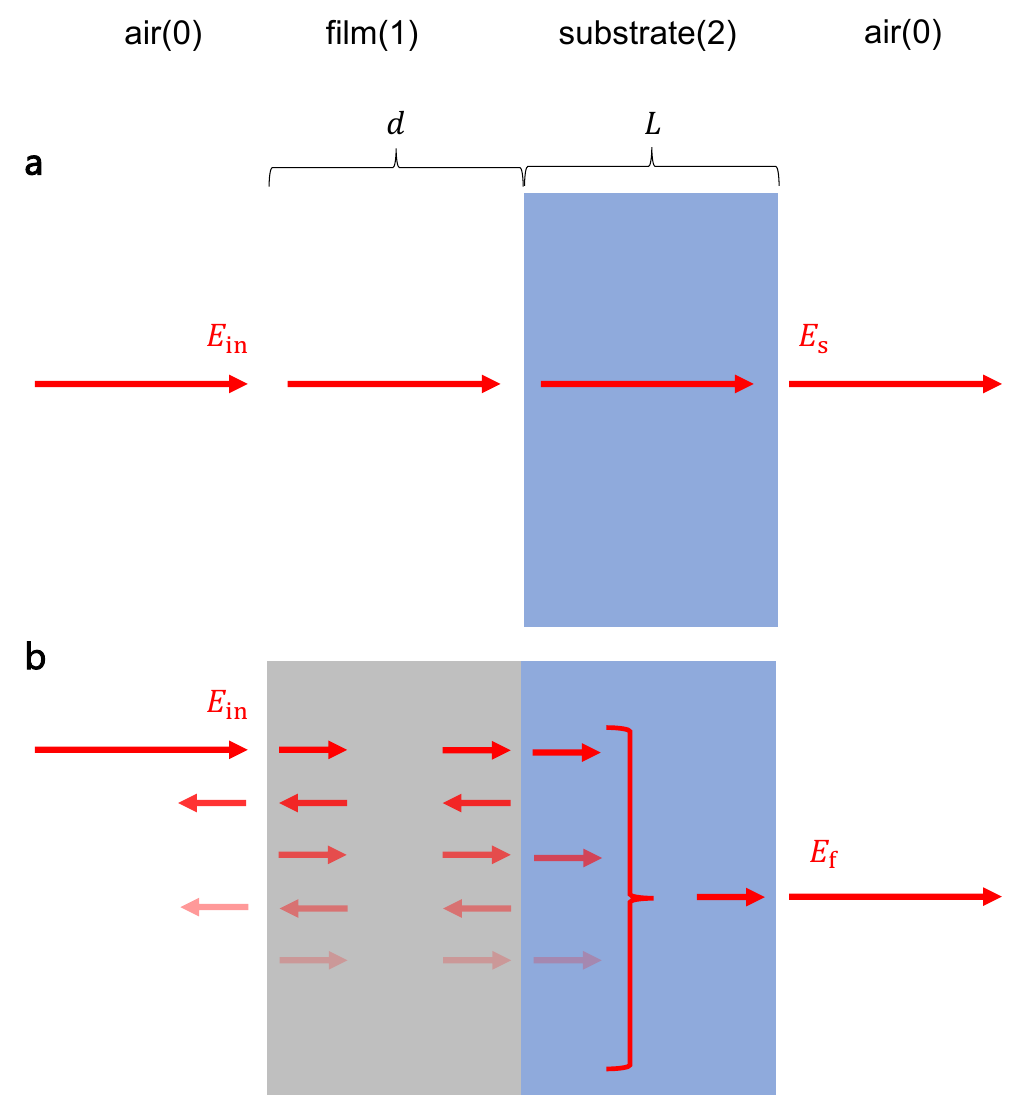}
\caption{Schematic picture of the pulse propagation {\bf a} with and {\bf b} without the FeSe film. The multiple reflections in the substrate are ignored.}\label{fig:S6}
\end{figure}

\begin{figure}[h]
\centering
\includegraphics[width=0.9\textwidth]{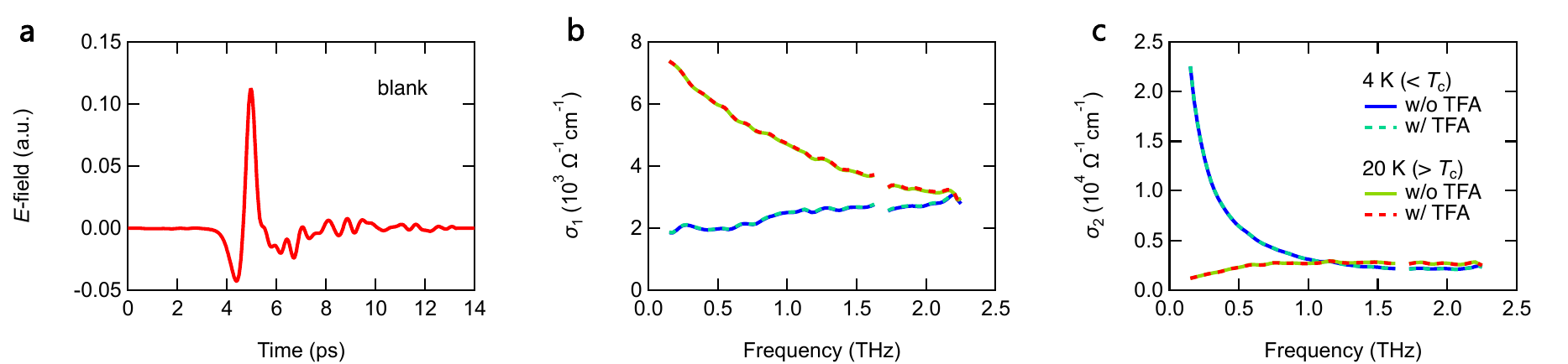}
\caption{{\bf a} The waveform of the electric field component of the THz probe pulse used for the optical conductivity measurements. This waveform was measured with placing a blank aperture of $6\, \text{mm}$ in diameter at the sample position. {\bf b} The real and {\bf c} the imaginary part of optical conductivity of FeSe sample, which is not exactly the same one used in the main text but grown in the same conditions, calculated with (dashed curve) and without (solid curve) the thin film approximation (TFA). The missing spectral range between $1.6\, \text{THz}$ to $1.7\, \text{THz}$ is caused by the residual water vapor absorption.}\label{fig:S7}
\end{figure}

In this section, we describe the thin-film approximation to extract the optical conductivity. To determine the optical conductivity, we measured the waveform of the transmitted pulse through only a substrate (Fig. \ref{fig:S6}(a)) and a substrate with a film (Fig. \ref{fig:S6}(b)). In the following analysis, we neglect the multiple reflection in the substrate because a nearly single-cycle THz pulse is used for the optical conductivity measurement which enables the elimination of the reflected pulses in the time domain. (Fig. \ref{fig:S7}(a))

The relationship between $E_{s}$, which is the electric field of the transmitted pulse through only a substrate just to the right of the substrate/air interface and $E_{in}$, which is the electric field of the incident pulse at the position corresponding to the film/substrate interface, is described by the following Fresnel equation:

\begin{equation}
    \label{eq:Es}
    E_{s} = E_{in}\frac{4n_{s}}{(n_{s}+1)^{2}}\exp{\left(i\omega\frac{n_{f}d+n_{s}L}{c} \right)}
\end{equation}

The relationship between $E_{f}$, which is the electric field of the transmitted pulse through the substrate with the film just to the right of the substrate/air interface, and $E_{in}$ is described by the Fresnel equation:

\begin{equation}
    \label{eq:Ef}
    \begin{aligned}
        E_{f} &= E_{in}\frac{8n_{f}n_{s}f_{f}}{(n_{f}+1)(n_{f}+n_{s})(n_{s}+1)}\exp{\left(i\omega\frac{n_{f}d+n_{s}L}{c} \right)} \\
        &= E_{in}\frac{8n_{f}n_{s}}{(n_{f}+1)(n_{f}+n_{s})(n_{s}+1)}\exp{\left(i\omega\frac{n_{f}d+n_{s}L}{c} \right)}\left[1-\frac{(n_{f}-n_{s})(n_{f}-1}{(n_{f}+n_{s})(n_{f}+1)}\exp{\left(i\omega\frac{2n_{f}d}{c}\right)}\right]^{-1}
    \end{aligned}
\end{equation}

where $f_{f}=\left[1-\frac{(n_{f}-n_{s})(n_{f}-1}{(n_{f}+n_{s})(n_{f}+1)}\exp{\left(i\omega\frac{2n_{f}d}{c}\right)}\right]^{-1}$ describes the Fabry-Perot effect within the film. Using these two relationships, the relationship between $E_{s}$ and $E_{f}$ can be derived as follows:

\begin{equation}
    \label{eq:Ef_Es}
    E_{f} = E_{s}\frac{2n_{f}(n_{s}+1)}{(n_{f}+1)(n_{f}+n_{s})}\exp{\left(i\omega\frac{(n_{f}-1)d}{c} \right)}\left[1-\frac{(n_{f}-n_{s})(n_{f}-1}{(n_{f}+n_{s})(n_{f}+1)}\exp{\left(i\omega\frac{2n_{f}d}{c}\right)}\right]^{-1}.
\end{equation}

We can simplify the above relationship if the film is very thin ($|\omega n_{f} d_{f}/c| \ll 1$) and the refractive index of the film is very large ($|n_{f}| \gg n_{s} > 1$). Using this thin film approximation and the relationship between the optical conductivity $\sigma$ and the refractive index $n$, $\sigma = -i\epsilon_{0} \omega(n^{2}-1)$, the above relationship is simplified as

\begin{equation}
    \label{eq:Ef_Es_simple}
    E_{f} = E_{s}\frac{n_{s}+1}{n_{s}+1+Z_{0}d\sigma_{f}}
\end{equation}

where $\sigma_{f}$ is the optical conductivity of the film.

To confirm that the condition for the thin-film approximation is satisfied, we compare the conductivities of our FeSe film sample using the relationships derived with and without the thin-film approximation, as shown in Figs. \ref{fig:S7}(b) and (c). The results are almost identical, indicating that the thin-film approximation is valid for our sample.
 The simplified relationship between $E_{s}$ and $E_{f}$ can also be derived by regarding the film as air/substrate boundary.
 
\begin{figure}[h]
\centering
\includegraphics[width=0.4\textwidth]{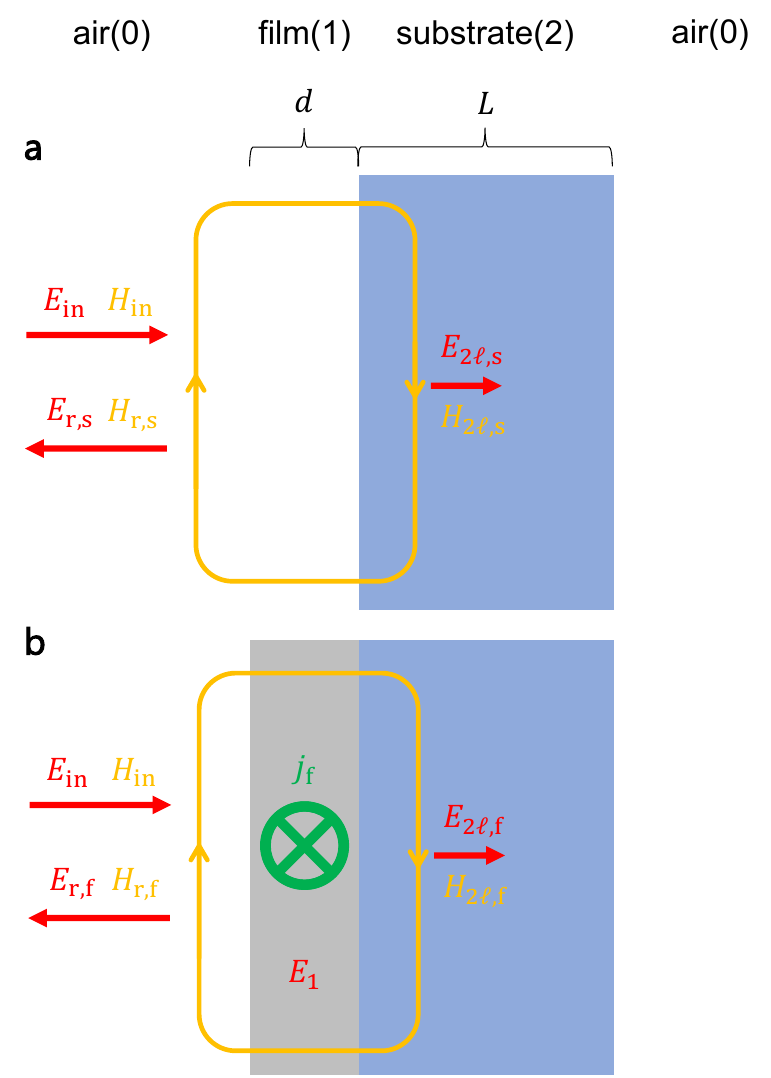}
\caption{Schematic picture of the boundary conditions at air/substrate interface {\bf a} without and {\bf b} with the FeSe film. The loops, where the boundary conditions considered are represented by orange loops. The green symbol represents the current inside the film induced by the incident pulse.}\label{fig:S8}
\end{figure}

 Without the film, the boundary conditions at the air/substrate boundary are described as follows: 

\begin{align}
    \label{eq:BC_E_wof}
    E_{in} + E_{r,s} = E_{2\ell,s} \\
    \label{eq:BC_H_wof}
    H_{in} - H_{r,s} = H_{2\ell,s}  
\end{align}

where, $E_{in}$ and $H_{in}$ represent the electric field and magnetic field of the incident pulse just to the left of the boundary, $E_{r,s}$ and $H_{r,s}$ represent the electric field and the magnetic field of the reflected pulse just to the left of the boundary and $E_{2\ell,s}$ and $H_{2\ell,s}$ represent the electric field and the magnetic field of the transmitted pulse just to the right of the boundary. (Fig. \ref{fig:S8}(a))

When the film exists, the boundary condition for the magnetic fields changes because the incident pulse induces the current $j_{f}$ at the boundary (in the film). Therefore, the boundary conditions with the film are described as follows. (Fig. \ref{fig:S8}(b))

\begin{align}
    \label{eq:BC_E_wf}
    E_{in} + E_{r,f} = E_{2\ell,f} \\
    \label{eq:BC_H_wf}
    H_{in} - H_{r,f} = H_{2\ell,f} + j_{f}d  
\end{align}

Using Ohm’s law, $j_{f}$ can be represented by $E_{1}$, an electric field inside the film, as $j_{f} = \sigma_{f} E_{1}$. Here, we assume $E_{1} = E_{2\ell,f}$. Then, from the above boundary conditions and the relationships $\mu_{0} H = nE/c$, we can derive the relationship between $E_{2\ell,s}$ and $E_{2\ell,f}$ as follows.

\begin{equation}
    \label{eq:Ef_Es_approx}
    E_{2\ell,f} = E_{2\ell,s}\frac{n_{s}+1}{n_{s}+1+Z_{0}d\sigma_{f}}
\end{equation}

Since the pulses propagate the same way after the left air/substrate boundary regardless of the existence of the film, the above relation is equal to that of $E_{f}$ and $E_{s}$ derived first.

\vskip\baselineskip

The fact that the optical conductivities derived using the simplified relation between $E_{f}$ and $E_{s}$ by the thin-film approximation are almost identical to those without thin film approximation, as shown in Figs. \ref{fig:S7}(b) and (c), indicates that the assumption we used in the second approach, $E_{1} = E_{2\ell,f}$, i.e. the electric field inside the film is equal to that of transmitted pulse just to the right of the film/substrate interface, is valid for our sample. The relationship $E_{1}=E_{2\ell,f}$ is useful to simplify not only the optical conductivity analysis but also the THG analysis as described in Supplementary Note 4.

\section*{Supplementary Note 6. Polarisation-resolved THG measurements}\label{sec:SI_PolDepMeasurements}

\begin{figure}[h]
\centering
\includegraphics[width=0.8\textwidth]{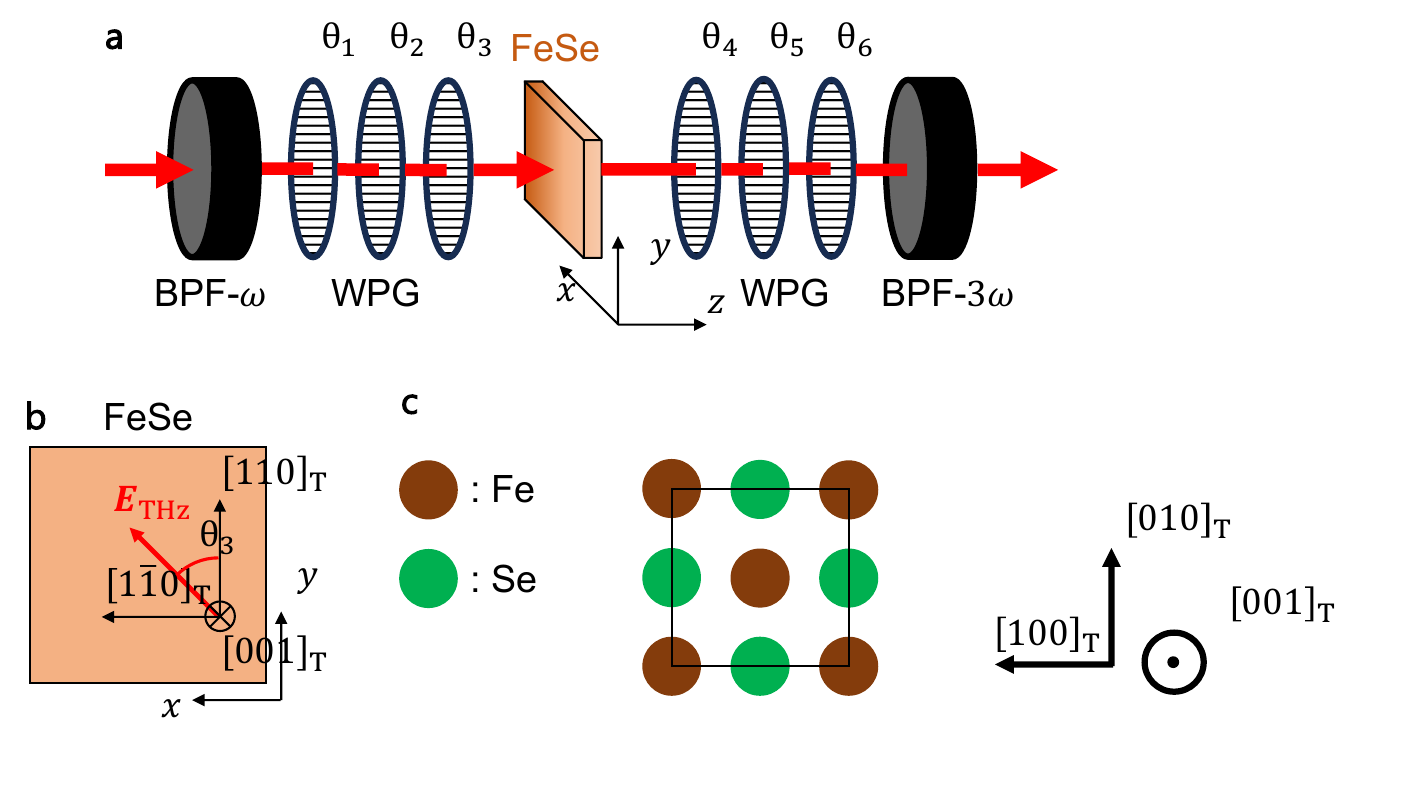}
\caption{{\bf a} Schematic setup for polarisation-resolved THG measurements. WGP: wire grid polarizer, BPF: band pass filter. {\bf b} The relationship between the crystal axis of the sample and the laboratory coordinate flame. {\bf c} Top view of the crystal structure of FeSe along with the crystal axis.}\label{fig:S9}
\end{figure}

To measure the polarisation dependence of the TH signal, several wire grid polarizers (WGPs) were used as illustrated in Fig. \ref{fig:S9}(a). In front of the sample, we placed three wire-grid polarizers WGP1, WGP2 and WGP3, whose respective angles from $y$-axis are $\theta_{1}$, $\theta_{2}$ and $\theta_{3}$. Note that all angles are defined as that from $y$-axis (see Fig. \ref{fig:S9}(b)). WGP1 was placed to ensure linear polarisation along the $y$-axis from the THz pulse generated in a $\text{LiNbO}_{3}$ crystal. Therefore, $\theta_{1} = 0\degree$ at all times. The polarisation state on the sample is determined solely by $\theta_{3}$. The angles $\theta_{3} = 0\degree$ and $45\degree$ correspond to the $[110]_{\text{T}}$ and $[100]_{\text{T}}$ directions, respectively, of the FeSe sample as shown in Fig. \ref{fig:S9}(b), where the subscript T represents the crystal axis orientation in the tetragonal phase. Because the field strength at the sample surface is factored in by $\cos\theta_{2}\cos{(\theta_{3}-\theta_{2})}$, we adjusted $\theta_{2}$ to keep the field-strength on the sample surface constant. Additional three polarizers WGP4, WGP5, and WGP6 were placed behind the sample with angles $\theta_{4}$, $\theta_{5}$ and $\theta_{6}$, respectively. WGP4 was set to $\theta_{4} = \theta_{3}$ for the detection of the TH signal polarized parallel to the incident pulse and set to $\theta_{4} = \theta_{3}+90\degree$ for the detection of the TH signal polarized perpendicular to the incident pulse. WPG6 was placed to ensure that the THz pulse incident to the EO crystal was linearly polarized along the y-axis. Therefore, $\theta_{6} = 0\degree$ at all times. The electric field along $\theta_{4}$ is then obtained by setting $\theta_{5} = +45\degree$ and $-45\degree$, using the following relation:

\begin{equation}
    \label{eq:E_PolDep}
    E = \sqrt{2}(E^{+}\cos{(\theta_{4}-45\degree)} - E^{-}\cos{(\theta_{4}-135\degree)}),
\end{equation}

where $E^{+}$ and $E^{-}$ represent the measured electric field with $\theta_{5}=+45\degree$ and $-45\degree$, respectively. 
The parallel ($E^{\parallel}$) and perpendicular ($E^{\perp}$) components of the TH signal with respect to the incident polarisation ($\theta_{3}$) are then taken by setting $\theta_{4} = \theta_{3}$ and $\theta_{4} = \theta_{3}+90\degree$, respectively.

To improve the signal-to-noise ratio, the polarisation dependence of $I_{\omega}$ was measured without $3\omega$-BPF.

\section*{Supplementary Note 7. THG associated with the Higgs mode for the case of anisotropic order parameter}\label{sec:SI_anisoGapTHGcalc}
In the case of full-gap $s$-wave SCs, the Higgs-mode resonance in the THG response appears at $\omega = \Delta(T)$ as we have seen for the case of NbN as detailed in the main text. In FeSe, by contrast, a resonance was identified far below the superconducting gaps $2\Delta$ (here the gap refers to the maximum one in the momentum space). However, the superconducting gaps of FeSe are strongly anisotropic \citesi{SI_doi:10.1126/science.aal1575,SI_PhysRevX.8.031033}, so that we should consider the effect of anisotropy on the Higgs- mode resonance for SCs such as FeSe. A similar problem has been studied theoretically, in particular for the case of $d$-wave SCs \citesi{SI_PhysRevB.101.184519,SI_PhysRevB.93.180507}, where the Higgs mode resonance was shown to appear slightly below the antinodal gap energy $2\Delta$ \citesi{SI_PhysRevB.101.184519}. 

Here, in order to examine the case of the $s+d$ type OP associated with the hole pocket, we perform the calculation in a single band model at $T$=0 considering the anisotropy of the $\Gamma$-pocket.
To represent the $s+d$ type OP in the hole pocket, we consider two $k$-dependent gap functions: one is the nodal gap of the form $\Delta_{k} \propto \Delta_0 (1 +\cos{2\theta_{k}} )$, which is estimated from ARPES measurement for the gap function of the hole band \citesi{SI_PhysRevX.8.031033}, and another is the nodeless gap of the form $\Delta_{k} \propto \Delta_0 (1 +0.65\cos{2\theta_{k}} )$, as estimated from ref.\citesi{SI_doi:10.1126/science.aal1575}. Then, we calculate the frequency dependence of the TH current associated with the Higgs mode oscillation using the model described in ref.\citesi{SI_PhysRevB.101.184519} to evaluate the impact of the gap anisotropy on the THG resonance originated from the Higgs mode. 

\begin{figure}[h]
\centering
\includegraphics[width=0.7\textwidth]{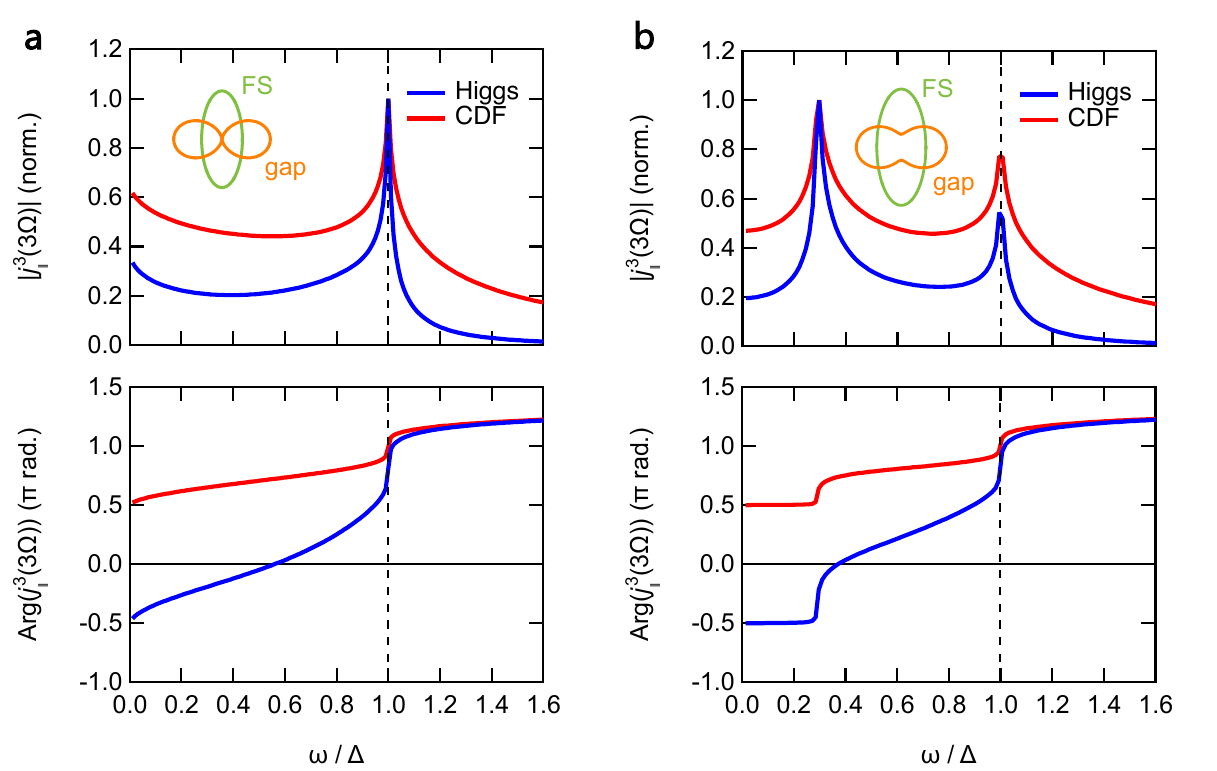}
    \caption{{\bf a,b} Calculated frequency dependence of the third-order nonlinear current originating from Higgs mode and CDF for the hole pocket with an s+d-like gap, $(1+\cos{2\theta})$ in {\bf a} and $(1+0.65\cos{2\theta})$ in {\bf b}. The blue curves represent the Higgs mode and the red curves the CDF mode, the inset represent the $s+d$-wave OP (orange) and the elliptic pocket at the $\Gamma$-point, assuming a single-band model. Each absolute values are normalized by their maximum values.}\label{fig:S10}
\end{figure}

The calculated results of the driving frequency dependence of the TH current are summarized in Figs. \ref{fig:S10}(a) and (b) upper panels, clearly showing that the TH signal enhances around $\omega \approx \Delta$ for both cases of the $s+d$ type OPs. More importantly, in addition to the absolute value of the TH current, the phase of the TH current shows a sudden change at $\omega = \Delta$ as shown in the lower panels of Figs. \ref{fig:S10}(a) and (b). We also calculate the TH current arising from the quasiparticle excitation term, termed charge density fluctuation (CDF) \citesi{SI_PhysRevB.93.180507}, showing that it also exhibits a resonance accompanied by a sudden phase change at the antinodal energy, $\omega = \Delta$. It should be noted here that in order to compare the relative magnitude between the Higgs and CDF contributions, it is crucial to include the paramagnetic coupling term taking into account the impurities effect, which is beyond the scope of the present work.

In contrast to this calculation, as mentioned in the main text, the phase shift of THG in FeSe was observed only at energies far below $2\Delta$, and no significant phase shift was observed around $2\Delta$. Furthermore, the phase jump was absent for the driving frequencies of $\omega=$0.3 THz and 0.5 THz(see Fig.2f in the main text) where $2\omega$ is obviously below the gaps $2\Delta$.
Therefore, the calculation results presented here suggest that the resonance of the THG in FeSe observed far below $2\Delta$ cannot be ascribed to the Higgs mode or to CDF, although a resonance peak appears in the proximity of the gap minimum for the nodeless case. 

We note here that the calculations presented here were performed for a model with a single Fermi surface, whereas FeSe is a multiband superconductor with multiple Fermi surfaces. As better shown in Supplementary Note 8, the material is a multiband system and all the equations are coupled together. Nevertheless, the absence of the THG resonance at $\omega = \Delta$ for either the hole pocket or the electron pocket implies the presence of a distinct low-energy mode, which is sharp and overwhelms all possible peak structures.

\section*{Supplementary Note 8.  Effective model for FeSe}
In this section, we propose an effective model for FeSe and discuss the various interactions and different pairing channels in the system.
Bulk FeSe is characterized by hole and electron Fermi pockets at the $\Gamma$ and $X/Y$ points of the Brillouin zone, as shown in Fig. 3(a) in the main text, similarly to other iron-pnictides but with a much smaller Fermi surface. 
Our sample is a thin film with a thickness of $70~{\rm nm}$ with a $T_c \approx 14 {\rm K}$, far from the single-layer case characterized by only electron pockets. We then consider the observed phenomenon as arising from a bulk response. 

In a 3-pocket model, the self-consistent gap equations read:

\begin{align*}
\Delta_{\vec{p}}^h & =-\int_{\vec{k}}\frac{\mathrm{d}^2 k}{(2 \pi)^2} \left[V_{\vec{p} \vec{k}}^{h e_X} \Delta_{\vec{k}}^{e_X} W_{\vec{k}}^{e_X}+V_{\vec{p} \vec{k}}^{h e_Y} \Delta_{\vec{k}}^{e_Y} W_{\vec{k}}^{e_Y}\right] \\
\Delta^{e_X} & =-\int_{\vec{k}}\frac{\mathrm{d}^2 k}{(2 \pi)^2} \left[V_{\vec{p} \vec{k}}^{h e_X} \Delta_{\vec{k}}^h W_{\vec{k}}^h\right], \\
\Delta^{e_Y} & =-\int_{\vec{k}}\frac{\mathrm{d}^2 k}{(2 \pi)^2} \left[V_{\vec{p} \vec{k}}^{h e_Y} \Delta_{\vec{k}}^h W_{\vec{k}}^h\right],
\end{align*}

where $W_{\vec{k}}^x=\frac{1}{2 E_{\vec{k}}^x} \tanh \frac{E_{\vec{k}}^x}{2 T} $, with $ x \in\left\{h, e_X, e_Y\right\}$.

Along the lines of ref. \citesi{SI_PhysRevLett.120.267001}, we consider the paring interaction $V_{\vec{p} \vec{k}}^{h e_i}$ to be of the form $ (A_i + B_i \cos{2\theta_h})$, with 
\begin{align}
\label{eq:fernandes}
    \begin{aligned}
   & A_{X,Y}=[U_s (1\mp 2\Phi_e)\mp U_d \Phi_h],\\
   & B_{X,Y}=[\pm U_d -2\Phi_e U_d],
   \end{aligned}
\end{align}
where  $\Phi_{h(e)}$ represents the nematic order parameter for the hole (electron) pocket and $U_{s(d)}$ denotes the coupling constant for the $s(d)$-wave pairing channels, respectively.
By linearizing the gap equations, it is possible to obtain the pairing interactions as obtained in ref.\citesi{SI_PhysRevLett.120.267001}, which we rewrite below:
\begin{align*}
\Delta_h\left(\boldsymbol{k}_h\right)=- & A^2\left[\left(A_X+B_X \cos 2 \theta_h\right) \int \frac{\mathrm{d}^2 k}{(2 \pi)^2} \frac{\tanh \left(\beta \epsilon_{X, \boldsymbol{k}} / 2\right)}{2 \epsilon_{X, \boldsymbol{k}}} \sin ^2 \theta_X \Delta_X\right. \\
& \left.+\left(A_Y+B_Y \cos 2 \theta_h\right) \int \frac{\mathrm{d}^2 k}{(2 \pi)^2} \frac{\tanh \left(\beta \epsilon_{Y, \boldsymbol{k}} / 2\right)}{2 \epsilon_{Y, \boldsymbol{k}}} \sin ^2 \theta_Y \Delta_Y\right] \\
\Delta_X\left(\boldsymbol{k}_X\right)=- & A^2 \sin ^2 \theta_X \int \frac{\mathrm{~d}^2 k}{(2 \pi)^2} \frac{\tanh \left(\beta \epsilon_{h, \boldsymbol{k}} / 2\right)}{2 \epsilon_{h, \boldsymbol{k}}}\left(A_X+B_X \cos 2 \theta_h\right) \Delta_{h, \boldsymbol{k}}, \\
\Delta_Y\left(\boldsymbol{k}_Y\right)=- & A^2 \sin ^2 \theta_Y \int \frac{\mathrm{~d}^2 k}{(2 \pi)^2} \frac{\tanh \left(\beta \epsilon_{h, \boldsymbol{k}} / 2\right)}{2 \epsilon_{h, \boldsymbol{k}}}\left(A_Y+B_Y \cos 2 \theta_h\right) \Delta_{h, \boldsymbol{k}}.
\end{align*}
In the absence of nematicity, the interactions between the hole and two electron pockets are symmetric whereas, in the nematic phase, as shown in ref.\citesi{SI_PhysRevLett.120.267001}, the situation changes and the interaction gets renormalized in such a way that the effective interaction between along $\Gamma X$ is stronger than $\Gamma Y$ \citesi{SI_Benfatto_2018} (see Fig. 3(a) in the main text). This is in accordance with the idea that the gap structures on the hole and electron pocket at X behave as if they were determined only by the distribution of the $d_{yz}$ orbital weight; see ref. \citesi{SI_PhysRevB.98.180503}. 

We parametrize the gap on the hole band as $\Delta_h=\Delta_1 + \Delta_2 \cos{2\theta_h}$ and the gaps on the electron pockets as $\Delta_X=\Delta_3 \sin^2 \theta_X, \Delta_Y=\Delta_4 \sin^2 \theta_Y$ and write the effective gap equations with retaining the $C_2$ symmetric band dispersion in the particle-particle bubbles $\Xi_e,\Xi_h$. 
In particular :
\begin{align*}
    &\Xi_e=\Xi_X=\Xi_Y=\int \frac{d^2k}{(2\pi)^2} \frac{\mathrm{tanh}(\beta \epsilon_{k,X/Y}/2)}{2\epsilon_{X/Y,k}}\sin^4{\theta_{X/Y}},\\
      &\Xi_{h,j}=\int \frac{d^2k}{(2\pi)^2} \frac{\mathrm{tanh}(\beta \epsilon_{k,h}/2)}{2\epsilon_{h,k}}(\cos{2\theta_{h}})^j  ~\text{with } j=0,1,2.
\end{align*}
By substituting the relations in \eqref{eq:fernandes} and expanding to the first order in $\Phi_e,\Phi_h$ we obtain: 
\begin{equation}
\label{eq:gaps}
\begin{aligned}
\binom{\Delta_1}{\Delta_2}\propto  \Xi^{e}\Xi^h\left(\begin{array}{cc}
V_s & W\\
W'& V_d
\end{array}\right)\binom{\Delta_1}{\Delta_2},
\end{aligned}
\end{equation}
which is equivalent to the equation in the main text. Here, $V_s$ and $V_d$ represent effective interactions reflecting the $s$- and $d$- wave pairing instabilities, respectively, and $W, W'$ are the mixing of the two channels.

Within this simplified picture, we can solve the gap equation by diagonalizing $\hat{V}$. The eigenchannel of the pairing interaction corresponding to the ground state of the system is then represented as the sign-preserving s-wave-dominated $\Delta(\theta_h)=\Delta_h (1+r \cos{2\theta_h})$, where we introduce a parameter $r$ representing the superconducting gap anisotropy.
We obtain $r\approx 0.5$ from the values of interactions as we describe in the following.
The other solution $\Delta(\theta_h)=\Delta^{'}_h (r'+ \cos{2\theta_h})$, with $r'\approx -0.5$, represents the subdominant pairing channel, which is sign changing and thus dominated by the $d$-wave pairing. The $\pm$ sign in the form factors is crucial with respect to the distorted $C_2$ Fermi surface. In fact, the solution with the sign $+$ represents the case of gap minima perpendicular to the direction of the Fermi surface minima, and the solution with the sign $-$ represents the case of gap minima along the direction of the Fermi surface minima. 
The mode associated to the subdominant pairing channel is then viewed as a Bardasis-Schrieffer mode, meaning a sharp in-gap mode associated with the formation of a bound state of the two electrons of a broken Cooper pair in correspondence with the subdominant pairing instability. 
The frequency associated to the mode can be expressed in terms of the interactions in the eigenchannels:
\[
\omega_{BS}\propto\frac{1}{g_l}-\frac{1}{g_0} \xrightarrow{ }\begin{cases} \omega_{BS} \approx 2\Delta_{\mathrm{min}} & \text { if } g_l \ll g_0 \\ \omega_{BS}\ll2\Delta_{\mathrm{min}} &\text { if } g_l \approx g_0\end{cases}
\]
where $g_0$ is the attractive interaction in the dominant pairing channel and $g_l$ the interaction in the subdominant channel, see refs.\citesi{SI_PhysRevB.92.094506,SI_Sun_2020, SI_Wan_2022,SI_PhysRevB.103.024519} for more details.

To obtain an estimate, we summarize in Table \ref{tab:para} the parameters obtained by considering an elliptic Fermi surface to model the hole pocket with the effective mass of $m_x=2.37m_y$ and adopting $U_s=U_d=U$, $\Phi_h=0.3,\Phi_e=-0.1$ as in ref.\citesi{SI_Huang_2018}.
From these parameters we estimate $g_l\approx g_0/3$, which leads to a Bardasis-Schrieffer mode right below the gap minima (with the above parameters, the estimate is roughly $\omega_{BS}\approx0.9 (2\Delta_{\mathrm{min}})$).
This observation is consistent with the energy of the mode observed experimentally, which lies in close proximity to the gap minima of the hole gap $2\Delta_{\mathrm{min}}\approx0.8{\rm meV}$ as measured experimentally in ref.\citesi{SI_doi:10.1126/science.aal1575} \\[1em]
\begin{center}
\begin{table}[h]
\begin{tabular}{c|c|c|c|c|c|c|c}
\hline
$V_s(U^2)$ & $V_d(U^2)$&$W(U^2)$&$W'(U^2)$&$g_0(U^2)$&$g_l(U^2)$&r&r'\\
\hline
2&1&0.6&0.7&2&0.7&0.5&-0.5\\
\hline
\end{tabular}
\caption{Parameters for interactions and anisotropy adopted for the calculation rounded to one decimal place.}
\label{tab:para}
\end{table}
\end{center}

\section*{Supplementary Note 9. Coupling with light}

This section presents a derivation of the coupling to light of the Bardasis-Schrieffer mode within the pseudospin formalism \citesi{SI_PhysRevB.104.144508}. We denote the ground-state pairing as $\Delta_0 f_{0k}$ and the subdominant pairing channel as $\Delta_1 f_{1k}$, where $f_{0k}=1+ r\cos{2\theta_h}$ and $f_{1k}=\cos{2\theta_h}-r'$ represent the form factors of the corresponding pairing channel.
In order to perform the calculation, we consider a simplified model by retaining only the hole pocket, and we write a generalized BCS Hamiltonian with a ground state of the form $\Delta_k = \Delta_{0}f_{0,k}$, coupled to a vector potential which represents a spatially homogeneous laser field. We also assume that $\Delta_{0}f_{0,k}$ is in a region of vicinity to the subdominant pairing fluctuation. We then consider the self-consistent gap equations $\Delta_l=-V_l \sum_k \gamma_{kl} <c_{-\mathbf{k} \downarrow}c_{\mathbf{k} \uparrow}>$, where $l=0,1$.

Following the steps outlined in ref.\citesi{SI_M_ller_2019}, we introduce the Anderson's pseudospin,
\begin{equation}
\sigma_{\mathrm{k}}=\frac{1}{2} \left(\begin{array}{c}
c_{\mathbf{k} \uparrow}^{\dagger} c_{-\mathbf{k} \downarrow}
\end{array}\right) \tau  \left(\begin{array}{c}
c_{\mathbf{k} \uparrow} \\
c_{-\mathbf{k} \downarrow}^{\dagger}
\end{array}\right),
\end{equation}
and we write down the associated Bloch equation of motion. 
The equation of motion of the Anderson pseudospins with respect to the BCS Hamiltonian then describes a precession in the $\mathbf{b}_{\mathbf{k}}$ field

\begin{equation}
 \partial_t \sigma_{\mathbf{k}}(t)=\mathrm{i}\left[H_{\mathrm{BCS}}(t), \sigma_{\mathbf{k}}(t)\right]=\mathbf{b}_{\mathbf{k}}(t) \times \sigma_{\mathbf{k}}(t)   
\label{eq:bloch}
\end{equation}
with
\begin{equation}
\mathbf{b}_{\mathbf{k}}=\left(-2 \Delta^{\prime} , 2 \Delta^{"}, \epsilon_{\mathbf{k}-e\mathbf{A}} + \epsilon_{\mathbf{k}+e\mathbf{A}}  \right).   
\end{equation}
We then linearize the Bloch equations with respect to deviation from equilibrium within the approximation of a small laser intensity
\begin{equation}
  \begin{aligned}
\left\langle\sigma_{\boldsymbol{k}}^x\right\rangle(t) & =\left\langle\sigma_{\boldsymbol{k}}^x\right\rangle^{\mathrm{eq}}+\delta\sigma_{\boldsymbol{k}}^x(t), \\
\left\langle\sigma_{\boldsymbol{k}}^y\right\rangle(t) & =\left\langle\sigma_{\boldsymbol{k}}^y\right\rangle^{\mathrm{eq}}+\delta\sigma_{\boldsymbol{k}}^y(t), \\
\left\langle\sigma_{\boldsymbol{k}}^z\right\rangle(t) & =\left\langle\sigma_{\boldsymbol{k}}^y\right\rangle^{\mathrm{eq}}+\delta\sigma_{\boldsymbol{k}}^z(t),
\end{aligned}  
\end{equation}
where the effect of the laser field enters the equation via $\delta b^{z}_k(t) = b^{z}_k(t) - \epsilon_k \approx \frac{e^2}{2} \sum_{ij}(\partial_{k_i}\partial_{k_j} \epsilon_k ) A_i(t)A_j(t) $.\\
From the linearized equations of motion we derive solutions for the real and imaginary components of the gap in the form
 \begin{equation}
\left(\begin{array}{c}
\delta\Delta^{'}_{\mathbf{k},l} \\
\delta\Delta^{''}_{\mathbf{k},l}
\end{array}\right) \propto e^2 A^2(\omega)\left(\begin{array}{c}
\Delta^{'}_{\mathbf{k},l} \\
\Delta^{''}_{\mathbf{k},l}
\end{array}\right)
 \end{equation}

from which we compute the spectra associated with the collective mode of the overall gap excited by a linearly polarized laser field.

\subsection*{Linearized equation of motions}

Here we present the linearized equations of motion with respect to the deviation from equilibrium in the frequency domain at $T$=0. The linearization of eq.\eqref{eq:bloch} can be written in a matrix form as

 \begin{equation}
 \label{eq:matrix}
 \begin{aligned}
\left(\begin{array}{c}
\delta\sigma^{x}_{\mathbf{k}}(\omega) \\
\delta\sigma^{y}_{\mathbf{k}}(\omega)\\
\delta\sigma^{z}_{\mathbf{k}}(\omega)
\end{array}\right) = \frac{1}{E^{eq}_k((2E^{eq}_k)^2 - \omega^2)}  \left(\begin{array}{ccc}
2\epsilon_k^2 & i\omega\epsilon_k & -2\epsilon_k\Delta^{eq}_k \\
 -i\omega\epsilon_k & 2(\Delta^{eq}_k)^2 + 2\epsilon_k^2 & i\omega\Delta^{eq}_k  \\
2\Delta^{eq}_k \epsilon_k &  i\omega\Delta^{eq}_k & -2 (\Delta^{eq}_k)^2
\end{array}\right) \left(\begin{array}{c}
\delta\Delta^{'}_{\mathbf{k}}(\omega) \\
\delta\Delta^{"}_{\mathbf{k}}(\omega)\\
\delta b^{z}_{\mathbf{k}}
\end{array}\right)
\end{aligned}
\end{equation}
using $\delta \sigma_k (\omega) = \sigma_k - \sigma^{eq}_k$ and $ \delta\Delta^{'}_k(\omega) = \delta\Delta^{'}_{l}f_{l,k}(\omega)$, $ \delta\Delta^{"}_k(\omega) = \delta\Delta^{"}_{l}f_{l,k}(\omega)$ and the effect of the laser, $\delta b_{\gamma k}^z(t)=b_{\gamma k}^z(t)-\epsilon_{\gamma k} \simeq$ $\left(e^2 / 2\right) \sum_{i j}\left(\partial_{k_i} \partial_{k_j} \epsilon_{\gamma k}\right) A_i(t) A_j(t)$. Inserting the linearized equations back in eq.\eqref{eq:matrix} and solving self-consistently we obtain the following equations for the fluctuations:
 \begin{equation}
\hat{\chi}\left(\begin{array}{c}
\delta\Delta^{'}_{l} \\
\delta\Delta^{"}_{l}
\end{array}\right) =  \frac{1}{2} A^2 e^2 \left(\begin{array}{c}
\lambda_l^{'}   \\
\lambda_l^{''}
\end{array}\right)\label{eq:matrix_supp}
 \end{equation}

 with

$$
\hat{\chi}=\begin{aligned}
\left(\begin{array}{cc}
\chi_{11}& \chi_{12}\\
\chi_{21}& \chi_{22}
\end{array} \right) 
\end{aligned}
$$
\begin{equation}
    \begin{aligned}
       &\chi_{11}=1-V_l \sum_k \frac{ 2 \epsilon_k^2  f^{2}_{l,k}}{E_k \left(4 E^{2}_k-\omega^2\right)} ,\\
       &\chi_{12}=(\chi_{21})^{*}=-2 V_l \sum_k \frac{ \epsilon_k i \omega  f^{2}_{l,k}}{E_k \left(4 E^{2}_k-\omega^2\right)},\\
       &\chi_{22}= 1-V_l \sum_k \frac{ 2 E_k^2  f^{2}_{l,k}}{E_k \left(4 E^{2}_k-\omega^2\right)}
    \end{aligned}
\end{equation}

and

\begin{align*}
&\lambda_l^{'}= V_l \sum_k  \frac{2  f_{kl} \Delta_{0k}}{E_k \left(4 E^{2}_k-\omega^2\right)}\frac{\partial^2 \epsilon_k}{\partial k^{2}_x} \epsilon_{k},\\
& \lambda_l ^{''}=V_l \sum_k  \frac{2 i \omega f_{kl} \Delta_{0k}}{E_k \left(4 E^{2}_k-\omega^2\right)}\frac{\partial^2 \epsilon_k}{\partial k^{2}_x} .
\end{align*}

These functions contain the dependence on the bands parameters, which is considered by expanding the band structure as follows \citesi{SI_PhysRevB.95.104503}:
\begin{equation}
\sum_k \delta\left(\epsilon-\epsilon_{ k}\right) \frac{\partial^2 \epsilon_{k}}{\partial k_i^2}=D\left(c_{0}+c_{1} \epsilon+c_{2} \epsilon^2 \cdots\right)    
\end{equation}
where $i=x,y$. Since we are interested in the low energy response, the relevant coefficients are going to be $c_0$ and $c_1$, representing the inverse of the effective mass and the nonparabolicity of the band, respectively.
So we can approximate:

\begin{align*}
&\lambda_l^{'}= V_l \sum_k  \frac{2  f_{kl} \Delta_{0k}}{E_k \left(4 E^{2}_k-\omega^2\right)}c_1,\\
& \lambda_l ^{''}=V_l \sum_k  \frac{2 i \omega f_{kl} \Delta_{0k}}{E_k \left(4 E^{2}_k-\omega^2\right)} c_0.
\end{align*}

By looking at eq.\eqref{eq:matrix_supp} above we can notice that while the amplitude fluctuations couple to the vector potential via $\left(e^2 A^2 / 2\right) \left(\partial^2_{k_x} \epsilon_{\gamma k}\right) \epsilon_{\gamma k} \propto c_1$, such that the response is expected to be weak in the clean limit and in particular 0 for the parabolic band case; the phase fluctuations couple via $\left(e^2 A^2 / 2\right) \left(\partial^2_{k_x} \epsilon_{\gamma k}\right) \propto c_0$, as in the case of a Bardasis-Schrieffer/ Leggett mode, the response is then expected to be strong.\\

The subdominant pairing channel can have two possible directions of fluctuations, either orthogonal to $\Delta_0$ on the complex plane along the imaginary axis, or parallel to $\Delta_0$. The imaginary fluctuation corresponds to the Bardasis-Schrieffer mode, while the 'real' one does not correspond to a pole; see ref.\citesi{SI_Sun_2020}. 

\subsection*{Polarisation dependence}
The resonance with the Bardasis-Schrieffer mode can be observed experimentally in the transmitted light field as discussed in ref.\citesi{SI_M_ller_2019}. The nonlinear coupling of the vector potential to the collective mode leads to higher-harmonic generation, where the lowest non-vanishing order is the third-harmonic generation. Its intensity $I^{\mathrm{THG}}$, which is proportional to the squared amplitude of the induced third-order current $j^{(3)}(3 \Omega)$
$$
I^{\mathrm{THG}} \propto\left|j^{(3)}(3 \Omega)\right|^2
$$
shows the same resonance as the propagator. In the following, we calculate the contribution to the THG of the Higgs mode of the system, the BS mode, and the charge density fluctuations, i.e. single-particle excitations resonating at the pair breaking energy $2 \Delta$.  The calculation is done in a simplified single-pocket picture where we consider only the hole pocket and its gap as being described by the following BCS self-consistent gap equation:
\begin{equation}
    \Delta_0=V_0\int_{-\epsilon_c}^{\epsilon_c} d\theta f_0(\theta) \frac{\Delta_0 f_0(\theta)}{2 E_k},
\end{equation}

where $f_0(\theta)=1+r\cos(2\theta)$ and $E_k$ contains the dispersion of the hole pocket that we consider to be elliptical, $\epsilon_k=\frac{k_x^2}{2m_x}+\frac{k_y^2}{2m_y}$ with $m_x\approx2m_y$, closely following the treatment in ref.\citesi{SI_PhysRevB.101.184519}. The aim of this calculation is to show the possible polarisation dependence arising from such an anisotropic form factor in a system with $C_2$ symmetry. 

Driving the superconductor periodically will induce an electric current 
$$
j(t)=e \sum_k v_{k-c A(t)}\left\langle n_k\right\rangle(t)
$$
with the group velocity $v_k=\nabla \epsilon_k$ and the charge density
$$
\left\langle n_k\right\rangle(t)=\left\langle c_{k \uparrow}^{\dagger} c_{k \uparrow}+c_{k\uparrow}^{\dagger} c_{k \downarrow}\right\rangle(t).
$$

To calculate the lowest-order response, we expand the velocity in $A_0$
$$
v_{k-e A(t)}=v_k-e \sum_j A_j(t) \partial_j v_k+\mathcal{O}\left(A_0^2\right).
$$

The charge density can be expressed by the $z$-component of the pseudospin and we obtain for the current
$$
j(t)=j^{(0)}(t)+j^{(1)}(t)+j^{(3)}(t)
$$
where
$$
\begin{aligned}
j^{(0)}(t) & =e \sum_k v_k\left\langle n_k\right\rangle(t), \\
j^{(1)}(t) & =-2 e^2 \sum_k \sum_j A_j(t) \partial_j, v_k\left(\left\langle\sigma_k^2\right\rangle(0)+\frac{1}{2}\right), \\
j^{(3)}(t) & =-2 e^2 \sum_k \sum_j A_j(t) \partial_j v_k z_k(t).
\end{aligned}
$$

The first term \( j^{(0)}(t) \) vanishes due to parity. The second term \( j^{(1)}(t) \) represents the current that oscillates at the driving frequency \( \Omega \). The third term \( j^{(3)}(t) \), proportional to \( A_j(t) z_k(t) \), represents the higher-harmonic generation in the lowest-order,  \( 3 \Omega \), since \( z_k(t) \) oscillates at \( 2 \Omega \). The induced current at any angle relative to the vector potential's polarisation can be decomposed into parallel and perpendicular components, which we calculate separately by inserting the vector potential expression and expanding its components.

\[
j_{\| \perp}^{(3)}(t)=j^{(3)}(t) \cdot \hat{e}_A^{\| \perp}=-2 e^2 A_0(t) \sum_k D_{c_k}^{\| \perp}(\chi) z_k(t),
\]
with the directions parallel $\hat{e}_A^{\|}=(\cos (\chi), \quad \sin (\chi))^{\top}$ and perpendicular $\hat{e}_A^{\perp}=(\sin (\chi), \quad-\cos (\chi))^{\top}$ to the light polarisation. For the terms containing the derivatives, we use 
\begin{align*}
&D_{\epsilon_k}^{||}(\chi)=\cos (\chi)^2\partial_{x x}^2 \epsilon_k+\sin({\chi})^2\partial_{y y}^2 \epsilon_k+2\cos({\chi})\sin({\chi})\partial_{x y}^2 \epsilon_k\\
& D_{\epsilon_k}^{\perp}(\chi)=\sin (\chi) \cos (\chi)\left(\partial_{x x}^2 \epsilon_k-\partial_{y y}^2 \epsilon_k\right)+\partial_{x y}^2 \epsilon_k\left(\sin ^2(\chi)-\cos ^2(\chi)\right).
\end{align*}

To obtain an expression for the spectrum, we perform a Fourier transform to frequency $\Omega$
$$
j_{\| \perp}^{(3)}(\Omega)=-2 e^2 \sum_k D_{c_k}^{\| \perp}(\chi) \frac{1}{\sqrt{2 \pi}} \int A_0\left(\Omega^{\prime}\right) z_k\left(\Omega-\Omega^{\prime}\right) \mathrm{d} \Omega^{\prime}.
$$

Using $A(t)=\sin{\omega t}$ for the vector potential, the convolution can be evaluated, and the current at $\Omega = 3 \omega$ reads
$$
j_{\parallel \perp}^{(3)}(3 \omega)=\mathrm{i} e^2 A_0 \sum_k D_{c_k}^{\parallel \perp}(\chi) z_k(2 \omega).
$$ We can see that the THG response $3 \omega$ is determined by the $2 \omega$ response of the system. For further analysis, we make use of the solution $z_k(\Omega)=z_k(s=i \Omega)$ of the linearized Bloch equations \eqref{eq:bloch}. The term $z_k(s)$ has three contributions, namely, one term proportional to the real part of the gap $\propto \delta \Delta^{\prime}_{0}(s)$, one term proportional to the imaginary part of the gap $\propto \delta \Delta^{\prime \prime}_{0}(s)$, one proportional to the BS mode $\propto \delta \Delta^{\prime \prime}_{l}(s)$ and one term proportional to driving $\gamma_k^{A^2}(s)$. Therefore, we write
$$
j_{\| \perp}^{(3)}(3 \omega)=j_{\| \perp}^{(3) \mathrm{H}}(3 \omega)+j_{\| \perp}^{(3) \mathrm{P}}(3 \omega)+j_{\| \perp}^{(3) \mathrm{BS}}(3 \omega)+j_{\| \perp}^{(3) \mathrm{C}}(3 \omega),
$$
where the indices H, P, BS,C stand for Higgs, phase, Bardasis-Schrieffer and charge, respectively. We insert the solution $z_k(s)$  into the expression for the current. To write the resulting expressions in a compact form, we define several quantities
$$
\begin{aligned}
& x^l_0(s)=V_l \sum_k \frac{f_{lk}^2}{E_k\left(4 E_k^2+s^2\right)} \tanh \left(\frac{E_k}{2 k_B T}\right), \\
& x^{0l}_1(s)=V_l \sum_k \frac{f_{0k}^2 f_{lk}^2 }{E_k\left(4 E_k^2+s^2\right)} \tanh \left(\frac{E_k}{2 k_B T}\right), \\
& x^l_2(s)=V_l \sum_k \frac{\epsilon_k f_{lk}^2}{E_k\left(4 E_k^2+s^2\right)} \tanh \left(\frac{E_k}{2 k_B T}\right), \\
& x^l_3(s)=V_l \sum_k \frac{\epsilon_k^2 f_{lk}^2}{E_k\left(4 E_k^2+s^2\right)} \tanh \left(\frac{E_k}{2 k_B T}\right),
\end{aligned}
$$

$$
\begin{aligned}
& x_{4|0l}^{\| \perp}(s)=V_l \sum_k \frac{D_{c_k}^{\| \perp} f_{0k}f_{lk}}{E_k\left(4 E_k^2+s^2\right)} \tanh \left(\frac{E_k}{2 k_B T}\right), \\
& x_{5|0l}^{\| \perp}(s)=V_l \sum_k \frac{\epsilon_k D_{\varepsilon_k}^{\| \perp} f_{0k}f_{lk}}{E_k\left(4 E_k^2+s^2\right)} \tanh \left(\frac{E_k}{2 k_B T}\right), \\
& x_{6}^{\| \perp}(s)=V_l \sum_k \frac{D_{c_k}^{\|} D_{c_k}^{\| \perp} f_{0k}^2}{E_k\left(4 E_k^2+s^2\right)} \tanh \left(\frac{E_k}{2 k_B T}\right) .
\end{aligned}
$$

It follows for the four contributions
$$
\begin{aligned}
& j_{\| \perp}^{(3) \mathrm{H}}(3 \omega)=\mathrm{i} e^2 A_0 \frac{1}{V_0} 2 \Delta_0 x_{5|00}^{\| \perp}(2 \mathrm{i} \omega) \delta \Delta_0^{\prime}(2 \mathrm{i} \omega), \\
& j_{\| \perp}^{(3) \mathrm{BS}}(3 \omega)=-\mathrm{i} e^2 A_0 \frac{1}{V_l} s \Delta_0  x_{4|0l}^{\| \perp}(2 i \omega) \delta \Delta_l^{\prime \prime}(2 \mathrm{i} \omega), \\
& j_{\| \perp}^{(3) \mathrm{P}}(3 \omega)=-\mathrm{i} e^2 A_0 \frac{1}{V_0} s \Delta_0 x_{4|00}^{\| \perp}(2 i \omega) \delta \Delta_0^{\prime \prime}(2 \mathrm{i} \omega), \\
& j_{\| \perp}^{(3) \mathrm{C}}(3 \omega)=-\mathrm{i} e^4 A_0^3 A^2(2 \mathrm{i} \omega) \frac{1}{V} \Delta_0^2 x_6^{\| \perp}(2 \mathrm{i} \omega).
\end{aligned}
$$

The real and imaginary parts of the gap expressed within the same formalism read
$$
\begin{aligned}
& \delta \Delta_l^{\prime}(s)=\Delta_0 e^2 A_0^2(s) \frac{s^2 x^l_2(s) x_{4|0l}^{\|}(s)+2 x_{5|0l}^{\|}(s)\left(2 \Delta^2 x^{0l}_1(s)+2 x^l_3(s)-1\right)}{2 s^2 x^l_2(s)^2+2\left(2 x^l_3(s)-1\right)\left(2 \Delta^2 x^{0l}_1(s)+2 x^l_3(s)-1\right)}, \\
& \delta \Delta_l^{\prime \prime}(s)=s \Delta_0 e^2 A_0^2(s) \frac{2 x^l_3(s) x_{4|0l}^{\|}(s)-x_{4|0l}^{\|}(s)-2 x^l_2(s)x_{5|0l}^{\|}(s)}{2 s^2 x^l_2(s)^2+2\left(2 x^l_3(s)-1\right)\left(2 \Delta^2 x^{0l}_1(s)+2 x^l_3(s)-1\right)}.
\end{aligned}
$$

\begin{figure}[h]
    \centering
    \includegraphics[scale=.45]{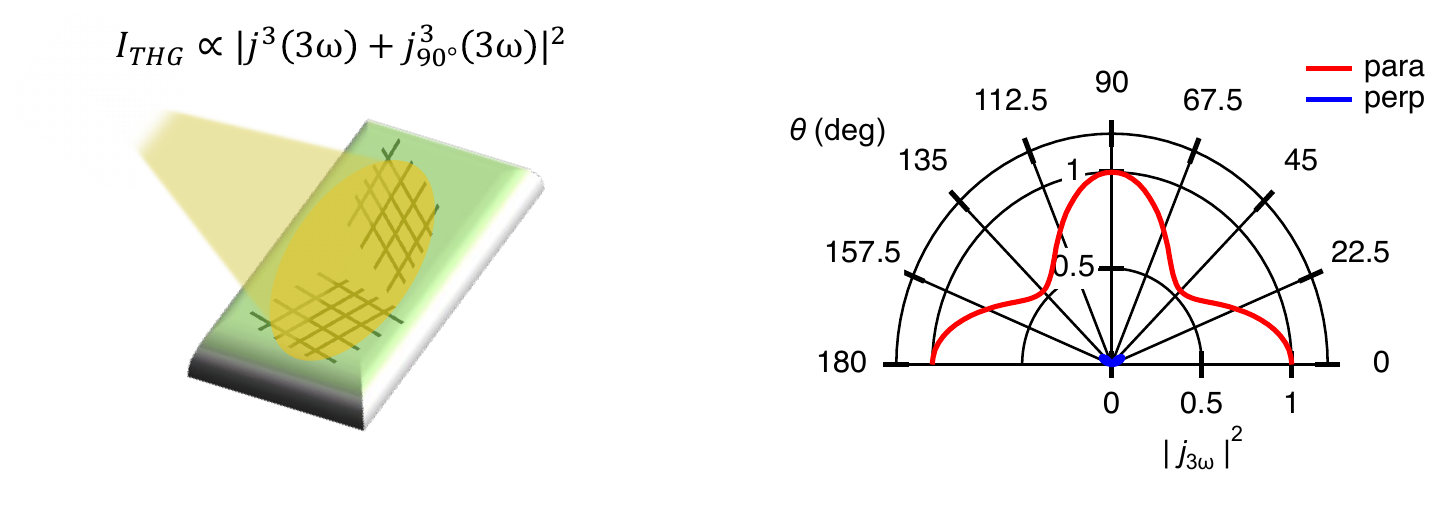}
    \caption{{\bf a} Cartoon representation of the laser spot on the sample, the signal detected is resulting from an average over the multi-domain structure of the sample. {\bf b} Calculated polarisation dependence for the Bardasis-Schrieffer mode in the parallel direction (red) and in the perpendicular direction (blue). The parameters used for the calculations on the hole pockets are: $c_{0x}=0.5, c_{0y}=0.2, c_{1x}=c_{1y}=-0.001$. The form factors for the ground state and for the metastable state are obtained from diagonalizing eq.\eqref{eq:gaps} and are respectively $f_{k0}=1+0.5\cos{2\phi}$ and $f_{kl}=-0.5+\cos{2\phi}$ , with $\Delta_0=1$, $V_0\approx3\times V_l$}
    \label{fig:PL_supp}
\end{figure}

Up to now, the expressions are exact within the linear regime. Now we proceed by putting $T=0$ and we consider the signal as the average on domains oriented $90\degree$ one respect to the other $
I^{\mathrm{THG}} \propto\left|j_{\chi}^{(3)}(3 \omega)+j_{\chi+\pi/2}^{(3)}(3 \omega)\right|^2
$, see Fig. \ref{fig:PL_supp}. Indeed, the polarisation dependence in our sample is measured by integrating over the multi-domain structure of FeSe.

\clearpage


\makeatletter
\let\thebibliographysi\thebibliography
\makeatother
\begin{thebibliographysi}{10}
\expandafter\ifx\csname url\endcsname\relax
  \def\url#1{\burl{#1}}\fi
\expandafter\ifx\csname urlprefix\endcsname\relax\def\urlprefix{URL }\fi
\providecommand{\bibinfo}[2]{#2}
\providecommand{\eprint}[2][]{\url{#2}}
\providecommand{\doi}[1]{\url{https://doi.org/#1}}
\bibcommenthead

\bibitem{SI_PhysRevB.90.121111}
\bibinfo{author}{Shimojima, T.} \emph{et~al.}
\newblock \bibinfo{title}{{Lifting of xz/yz orbital degeneracy at the structural transition in detwinned FeSe}}.
\newblock \emph{\bibinfo{journal}{Phys. Rev. B}} \textbf{\bibinfo{volume}{90}}, \bibinfo{pages}{121111} (\bibinfo{year}{2014}).

\bibitem{SI_Imai2017}
\bibinfo{author}{Imai, Y.} \emph{et~al.}
\newblock \bibinfo{title}{{Control of structural transition in $\text{FeSe}_{1\ensuremath{-}x}\text{Te}_{x}$ thin films by changing substrate materials}}.
\newblock \emph{\bibinfo{journal}{Sci. Rep.}} \textbf{\bibinfo{volume}{7}}, \bibinfo{pages}{46653} (\bibinfo{year}{2017}).

\bibitem{SI_10.1063/1.4826945}
\bibinfo{author}{Nabeshima, F.}, \bibinfo{author}{Imai, Y.}, \bibinfo{author}{Hanawa, M.}, \bibinfo{author}{Tsukada, I.} \& \bibinfo{author}{Maeda, A.}
\newblock \bibinfo{title}{{Enhancement of the superconducting transition temperature in FeSe epitaxial thin films by anisotropic compression}}.
\newblock \emph{\bibinfo{journal}{Appl. Phys. Lett.}} \textbf{\bibinfo{volume}{103}}, \bibinfo{pages}{172602} (\bibinfo{year}{2013}).

\bibitem{SI_doi:10.1126/science.aal1575}
\bibinfo{author}{Sprau, P.~O.} \emph{et~al.}
\newblock \bibinfo{title}{{Discovery of orbital-selective Cooper pairing in FeSe}}.
\newblock \emph{\bibinfo{journal}{Science}} \textbf{\bibinfo{volume}{357}}, \bibinfo{pages}{75--80} (\bibinfo{year}{2017}).

\bibitem{SI_PhysRevB.84.134522}
\bibinfo{author}{Schachinger, E.} \& \bibinfo{author}{Carbotte, J.~P.}
\newblock \bibinfo{title}{Finite-temperature signatures of gap anisotropy in optical conductivity of ferropnictides}.
\newblock \emph{\bibinfo{journal}{Phys. Rev. B}} \textbf{\bibinfo{volume}{84}}, \bibinfo{pages}{134522} (\bibinfo{year}{2011}).

\bibitem{SI_PhysRevB.100.035110}
\bibinfo{author}{Yoshikawa, N.} \emph{et~al.}
\newblock \bibinfo{title}{{Charge carrier dynamics of FeSe thin film investigated by terahertz magneto-optical spectroscopy}}.
\newblock \emph{\bibinfo{journal}{Phys. Rev. B}} \textbf{\bibinfo{volume}{100}}, \bibinfo{pages}{035110} (\bibinfo{year}{2019}).

\bibitem{SI_PhysRevX.8.031033}
\bibinfo{author}{Liu, D.} \emph{et~al.}
\newblock \bibinfo{title}{{Orbital origin of extremely anisotropic superconducting gap in nematic phase of FeSe superconductor}}.
\newblock \emph{\bibinfo{journal}{Phys. Rev. X}} \textbf{\bibinfo{volume}{8}}, \bibinfo{pages}{031033} (\bibinfo{year}{2018}).

\bibitem{SI_PhysRevB.101.184519}
\bibinfo{author}{Schwarz, L.} \& \bibinfo{author}{Manske, D.}
\newblock \bibinfo{title}{{Theory of driven Higgs oscillations and third-harmonic generation in unconventional superconductors}}.
\newblock \emph{\bibinfo{journal}{Phys. Rev. B}} \textbf{\bibinfo{volume}{101}}, \bibinfo{pages}{184519} (\bibinfo{year}{2020}).

\bibitem{SI_PhysRevB.93.180507}
\bibinfo{author}{Cea, T.}, \bibinfo{author}{Castellani, C.} \& \bibinfo{author}{Benfatto, L.}
\newblock \bibinfo{title}{{Nonlinear optical effects and third-harmonic generation in superconductors: Cooper pairs versus Higgs mode contribution}}.
\newblock \emph{\bibinfo{journal}{Phys. Rev. B}} \textbf{\bibinfo{volume}{93}}, \bibinfo{pages}{180507} (\bibinfo{year}{2016}).

\bibitem{SI_PhysRevLett.120.267001}
\bibinfo{author}{Kang, J.}, \bibinfo{author}{Fernandes, R.~M.} \& \bibinfo{author}{Chubukov, A.}
\newblock \bibinfo{title}{{Superconductivity in FeSe: the role of nematic order}}.
\newblock \emph{\bibinfo{journal}{Phys. Rev. Lett.}} \textbf{\bibinfo{volume}{120}}, \bibinfo{pages}{267001} (\bibinfo{year}{2018}).

\bibitem{SI_Benfatto_2018}
\bibinfo{author}{Benfatto, L.}, \bibinfo{author}{Valenzuela, B.} \& \bibinfo{author}{Fanfarillo, L.}
\newblock \bibinfo{title}{{Nematic pairing from orbital-selective spin fluctuations in FeSe}}.
\newblock \emph{\bibinfo{journal}{npj Quant. Mater.}} \textbf{\bibinfo{volume}{3}} (\bibinfo{year}{2018}).

\bibitem{SI_PhysRevB.98.180503}
\bibinfo{author}{Rhodes, L.~C.} \emph{et~al.}
\newblock \bibinfo{title}{{Scaling of the superconducting gap with orbital character in FeSe}}.
\newblock \emph{\bibinfo{journal}{Phys. Rev. B}} \textbf{\bibinfo{volume}{98}}, \bibinfo{pages}{180503} (\bibinfo{year}{2018}).

\bibitem{SI_PhysRevB.92.094506}
\bibinfo{author}{Maiti, S.} \& \bibinfo{author}{Hirschfeld, P.~J.}
\newblock \bibinfo{title}{{Collective modes in superconductors with competing $s$- and $d$-wave interactions}}.
\newblock \emph{\bibinfo{journal}{Phys. Rev. B}} \textbf{\bibinfo{volume}{92}}, \bibinfo{pages}{094506} (\bibinfo{year}{2015}).

\bibitem{SI_Sun_2020}
\bibinfo{author}{Sun, Z.}, \bibinfo{author}{Fogler, M.~M.}, \bibinfo{author}{Basov, D.~N.} \& \bibinfo{author}{Millis, A.~J.}
\newblock \bibinfo{title}{Collective modes and terahertz near-field response of superconductors}.
\newblock \emph{\bibinfo{journal}{Phys. Rev. Res.}} \textbf{\bibinfo{volume}{2}} (\bibinfo{year}{2020}).

\bibitem{SI_Wan_2022}
\bibinfo{author}{Wan, W.} \emph{et~al.}
\newblock \bibinfo{title}{{Observation of superconducting collective modes from competing pairing instabilities in single‐layer $\mathrm{NbSe}_{2}$}}.
\newblock \emph{\bibinfo{journal}{Adv. Mater.}} \textbf{\bibinfo{volume}{34}} (\bibinfo{year}{2022}).

\bibitem{SI_PhysRevB.103.024519}
\bibinfo{author}{M\"uller, M.~A.}, \bibinfo{author}{Volkov, P.~A.}, \bibinfo{author}{Paul, I.} \& \bibinfo{author}{Eremin, I.~M.}
\newblock \bibinfo{title}{{Interplay between nematicity and Bardasis-Schrieffer modes in the short-time dynamics of unconventional superconductors}}.
\newblock \emph{\bibinfo{journal}{Phys. Rev. B}} \textbf{\bibinfo{volume}{103}}, \bibinfo{pages}{024519} (\bibinfo{year}{2021}).

\bibitem{SI_Huang_2018}
\bibinfo{author}{Huang, W.}, \bibinfo{author}{Sigrist, M.} \& \bibinfo{author}{Weng, Z.-Y.}
\newblock \bibinfo{title}{{Identifying the dominant pairing interaction in high-Tc FeSe superconductors through Leggett modes}}.
\newblock \emph{\bibinfo{journal}{Phys. Rev. B}} \textbf{\bibinfo{volume}{97}} (\bibinfo{year}{2018}).

\bibitem{SI_PhysRevB.104.144508}
\bibinfo{author}{M\"uller, M.~A.} \& \bibinfo{author}{Eremin, I.~M.}
\newblock \bibinfo{title}{{Signatures of Bardasis-Schrieffer mode excitation in third-harmonic generated currents}}.
\newblock \emph{\bibinfo{journal}{Phys. Rev. B}} \textbf{\bibinfo{volume}{104}}, \bibinfo{pages}{144508} (\bibinfo{year}{2021}).

\bibitem{SI_M_ller_2019}
\bibinfo{author}{Müller, M.~A.}, \bibinfo{author}{Volkov, P.~A.}, \bibinfo{author}{Paul, I.} \& \bibinfo{author}{Eremin, I.~M.}
\newblock \bibinfo{title}{Collective modes in pumped unconventional superconductors with competing ground states}.
\newblock \emph{\bibinfo{journal}{Phys. Rev. B}} \textbf{\bibinfo{volume}{100}} (\bibinfo{year}{2019}).

\bibitem{SI_PhysRevB.95.104503}
\bibinfo{author}{Murotani, Y.}, \bibinfo{author}{Tsuji, N.} \& \bibinfo{author}{Aoki, H.}
\newblock \bibinfo{title}{{Theory of light-induced resonances with collective Higgs and Leggett modes in multiband superconductors}}.
\newblock \emph{\bibinfo{journal}{Phys. Rev. B}} \textbf{\bibinfo{volume}{95}}, \bibinfo{pages}{104503} (\bibinfo{year}{2017}).

\end{thebibliographysi}

\end{document}